\documentclass[aip,reprint]{revtex4-1}
\usepackage{graphicx} % Required for inserting images
\usepackage{dcolumn}% Align table columns on decimal point
\usepackage{bm}% bold math
\usepackage[utf8]{inputenc}
\usepackage[T1]{fontenc}
\usepackage{mathptmx}
\usepackage{etoolbox}
\usepackage[table,xcdraw]{xcolor}
\usepackage{subfig}
\usepackage{colortbl}
\usepackage{multirow}
\usepackage{caption}  % For subfigure environments (better control)
\usepackage{subcaption}
\usepackage{hyperref}
\hypersetup{
    colorlinks=true,
    citecolor=blue,
    linkcolor=blue,     
    }

\makeatletter
\def\@email#1#2{%
 \endgroup
 \patchcmd{\titleblock@produce}
  {\frontmatter@RRAPformat}
  {\frontmatter@RRAPformat{\produce@RRAP{*#1\href{mailto:#2}{#2}}}\frontmatter@RRAPformat}
  {}{}
}%
\makeatother

\begin{document}

% Use the \preprint command to place your local institutional report number 
% on the title page in preprint mode.
% Multiple \preprint commands are allowed.
%\preprint{AIP/123-QED}

\title[Review]{Addressing rotational motion on gravitational waves detectors. A review} %Title of paper

% repeat the \author .. \affiliation  etc. as needed
% \email, \thanks, \homepage, \altaffiliation all apply to the current author.
% Explanatory text should go in the []'s, 
% actual e-mail address or url should go in the {}'s for \email and \homepage.
% Please use the appropriate macro for the type of information

% \affiliation command applies to all authors since the last \affiliation command. 
% The \affiliation command should follow the other information.

\author{Chiara Di Fronzo}
\email[email: ]{chiara.difronzo@uwa.edu.au}
%\homepage[]{Your web page}
%\thanks{}
%\altaffiliation{}
\author{Jian Liu}
%\email[]{}
%\homepage[]{Your web page}
%\thanks{}
%\altaffiliation{}
\affiliation{The University of Western Australia, department of Physics, Maths and Computing, 35 Stirling Highway, Perth WA 6009, Australia}

% Collaboration name, if desired (requires use of superscriptaddress option in \documentclass). 
% \noaffiliation is required (may also be used with the \author command).
%\collaboration{}
%\noaffiliation

\date{\today}

\begin{abstract}
The rotational components of Earth's seismic motion are one of the major contributions to limit the sensitivity of terrestrial gravitational-wave detectors at frequencies below 10 Hz. The fundamental challenges lie in understanding the angular degrees of freedom of seismic motion and how they can be accurately measured. These are both crucial steps for developing an adequate control system to suppress seismic motion and maintaining resonance in the detector cavities. This review shows why the rotational ground motion limits the detectors' sensitivity and gives an overview of the technological achievements of the last decade on both the sensing and control systems sides. Perspectives on future developments in the field are also briefly illustrated.
\end{abstract}

% insert suggested PACS numbers in braces on next line

\maketitle %\maketitle must follow title, authors, abstract and \pacs

% Body of paper goes here. Use proper sectioning commands. 
% References should be done using the \cite, \ref, and \label commands
\section{Introduction}
Since the discovery of gravitational waves (GW) in 2015 \cite{Abbott2016}, detection rates and duty cycle increased over the years thanks to the technological improvements of the detectors \cite{Abbott2004, Pitkin2011, Riles2013, Allocca2020}, which targeted and addressed the noises affecting their sensitivities. These noises are characteristic of each of the main GW detectors operating around the world (Advanced LIGO, USA \cite{Aasi2015}; Advanced Virgo, Italy \cite{Acernese2015}; KAGRA, Japan \cite{Akutsu2019}). Noises depend on the location of the detector and their specific structures, shaping their sensitivities. However, the sensitivity curves show some common features, demonstrating that the detectors suffer from common or similar sources of noise. The one addressed in this review is seismic noise, which is affecting all the aforementioned ground-based GW detectors because of the fact that they are located on Earth and they are subjected to its ground motion.  The frequency bandwidth where the seismic noise shows its effect is referred to as the "low-frequency bandwidth", and it generally includes frequencies on and below 10 Hz.\\

\begin{figure}
    \centering
    \includegraphics[width=0.9\linewidth]{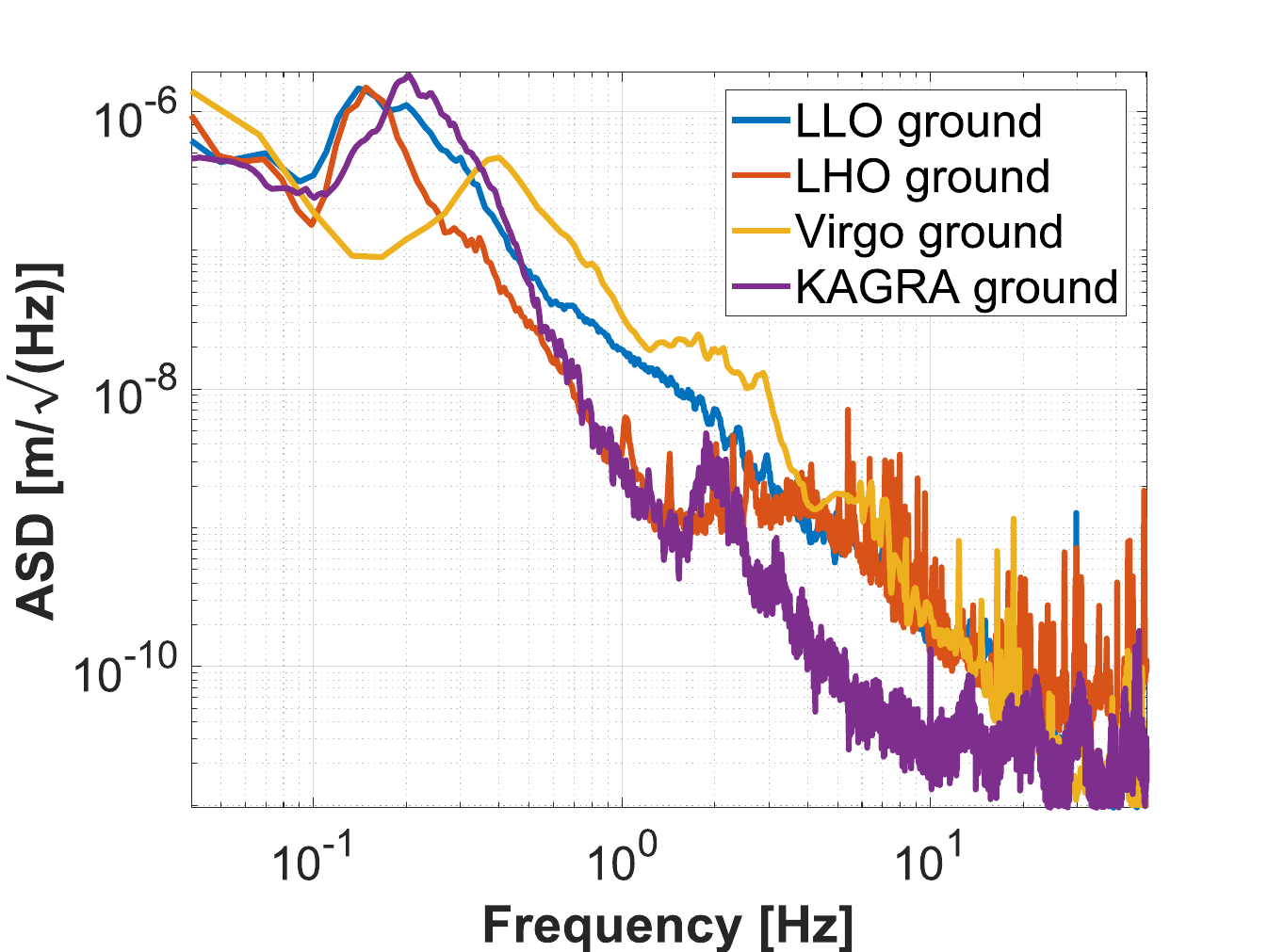}
    \caption{Seismic motion amplitude spectral densities of all 3 detectors compared to each other.}
    \label{fig:sei_all}
\end{figure}

Seismic noise is characterized by different features and sources, and it affects the detectors in different ways \cite{Daw2004,Beker2011,Accadia2012, Figura2022}. Most of the contributions come at the microseismic peak (0.1-0.5 Hz) due to the ocean waves exciting surface waves on the ground. At lower frequencies, the variation of gravity gradients due to mass motion on or below the ground induce a change in the density of the ground around the masses, affecting their stability (Newtonian Noise) \cite{Trozzo2022}. The ground motion occurs in all degrees of freedom (DOFs) (3DOF longitudinal and 3dof rotational). The one along the 3 rotational DOF, often referred to as "tilt", is the contribution addressed in this paper. Tilt affects not only the components of the detectors (up to, ultimately, the test masses) but also the sensors dedicated to detecting motions in other DOFs because it couples with other DOFs, providing the sensors with spurious output in their measurements and approximation in the control systems, due to the limitations in targeting and controlling this motion. This is an important contribution limiting the interferometer length and control systems from holding the positions of the optics accurately enough \cite{harms2020observation}, and it is observed as a steep curve in the ground spectra, generally below 10 Hz. Fig. \ref{fig:sei_all} shows an example of ground motion measured in all four sites. All the traces show a steep feature scaling as 1/$f^2$ from 10 Hz below. This is a characteristic feature of ground motion amplitude spectral densities of all detectors and it is a major limit to the detector sensitivities below 10 Hz. In this review we will explore why tilt couplings occur and how they are addressed.\\

Generally, rotational seismic motion is, historically, a widely investigated research field in seismology \cite{lee2009recent, kislov2021rotational, lee2009introduction}, and rotational ground motion components have been observed and studied during earthquakes \cite{igel2007broad} and volcanic eruptions \cite{eibl2022}, significantly improving the understanding of seismic motion in critical conditions. Substantial research progress has recently been done in understanding and processing 6DOF data \cite{yuan2020six, schmelzbach2018advances} via improving sensors and data analysis tools.\\
In this review, we describe how rotational motion is also important for gravitational-wave research, how it limits the detectors' sensitivities and what the state-of-the-art strategies to address this noise are. Sec. \ref{rot} A-D describes the tilt coupling effect. In Sec. \ref{angular} A-B, we describe the state-of-the-art of the sensing and control strategies; the possible strategies for the future generation detectors currently under study are briefly discussed in Sec. \ref{future}. A conclusive summary is presented in Sec. \ref{conclus}.
\section{Rotational motion on GW detectors}
\label{rot}
Ground-based detectors make use of vibration isolation techniques to keep the test masses stable within the detection requirements. Typically, the ground introduces a root-mean-square displacement (rms) of the order of magnitude of 1 $\mu$m. Detector cavities must be held at an rms of about 10$^{-14}$ m to ensure resonance and long duty cycles \cite{Abbott2004, Ross2020}. To achieve this goal, detectors are equipped with vibration isolation systems, which use passive, active, and a combination of active and passive strategies to reduce the seismic motion on the test masses and auxiliary optics.
Various sources of seismic noise are addressed by the vibration isolation systems of GW detectors, and in this review we focus on the motion in rotational degrees of freedom. Rotational (or tilt) motion occurs when the measured target rotates around one or more axes. This target can be the test masses or other optics of the detector, the inertial mass of an inertial sensor measuring longitudinal ground motion, the detector's platforms monitored and controlled to reduce vibrations, the pivot point of the detector's inverted pendula. We will see in this review this occurrences in detail. Table \ref{tab:rot} visually shows each rotational degree of freedom along the relative axis, using an optic as an example. In this review we will investigate its role on the ground-based detectors in three main occurrences: A) ground tilt motion affecting the active platforms where the optics are placed, and the seismometers dedicated to seismic motion sensing; B) tilt effect on 1) auxiliary optics cavities and 2) aLIGO test masses suspensions and C) tilt to length coupling effect induced by ground tilt motion onto inverted pendula suspensions of test masses on Virgo/KAGRA.

\begin{table}[h]
\centering
\resizebox{0.25\textwidth}{!}{
\begin{tabular}{c|c|c}
\rowcolor[HTML]{EFEFEF} 
\textbf{DoF} & \textbf{Name} & \cellcolor[HTML]{EFEFEF}                            \\ \cline{1-2}
$\theta_z$     & Yaw           & \cellcolor[HTML]{EFEFEF}                            \\
$\theta_y$     & Pitch         & \cellcolor[HTML]{EFEFEF}                            \\
$\theta_x$     & Roll          & \multirow{-4}{*}{\cellcolor[HTML]{EFEFEF} 
{\includegraphics[width=0.07\textwidth]{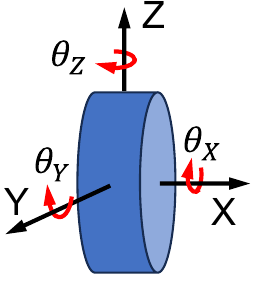}}
%{\includesvg[width=0.125\textwidth]{rotational_diagram.svg}}
}\\
\end{tabular}
}
\caption{Summary of different types of angular motion that affect the suspended test masses of a GW detector. Here X represents the direction of the optical axis, while Z represents the vertical suspension direction. }
\label{tab:rot}
\end{table}

\subsection{Rotational motion effect on active platforms and inertial sensors}
\label{tilt}
Detectors that make use of active vibration isolation, such as aLIGO, are affected by the angular degree of freedom of the ground motion at the level of their active platform, where test masses and auxiliary optics lie. \textit{Matichard et al. 2015} offer the most recent detailed overview of the aLIGO seismic isolation system \cite{Matichard2015}. It involves the use of an Internal Seismic Isolation platform (ISI) which is provided with sensors and actuators. The test masses and the beam splitter are suspended from these platforms and placed in vacuum inside a Basic Symmetric Chamber (BSC). The auxiliary optics lie on ISIs and the whole set is placed in vacuum in the Horizontal Access Module chambers (HAM). BSCs and HAMs designs differ from each other, with the BSC hosting two stages of ISI. The suspensions of auxiliary optics provide levels of passive isolation above 10 Hz. The ISI platforms reduce motion at a lower frequency ($\sim$ 0.1 Hz). A hydraulic system of attenuators (the Hydraulic External Pre-Isolator (HEPI)) and geophones provides isolation from ground motion. The devices installed to monitor and provide feedback are inertial and displacement sensors, which operate in vertical and horizontal directions. Some updates to the seismic system have been implemented in recent years, mainly to the control schemes \cite{Schwartz2020, DiFronzo2025} and more are planned to the hardware stage in anticipation of an increase of the test masses from 40 Kg to 100 Kg \cite{Bonilla2025}. In parallel, studies are ongoing to improve the seismic sensors and actuators of aLIGO \cite{Bonilla2025, Lantz2023}. On Virgo, inertial sensors are deployed on the suspended benches hosting the auxiliary optics and the injection beam \cite{acernese2015advanced}.\\
In general, the main challenge with ground tilt motion lies in directly measuring it because complete ground motion is not fully understood (in 6DOF) and all the rotational motions couple with the other degrees of freedom in most sensors measuring translational motion. Specifically, tilt motion couples with the horizontal translation motion of inertial sensors, and when it is extremely small ($<$10$^{-9}$ rad/$\sqrt{\rm Hz}$) it is challenging to decouple and measure, due to the nature of the sensor. In horizontal sensors, when the spring-mass system is tilted, an unwanted effect is induced at lower frequencies (10$^{-1}$ m/$\sqrt{\rm Hz}$), which pollutes the horizontal translation measurement.\\

\begin{figure}[h]
\subfloat[]{\includegraphics[scale=0.3]{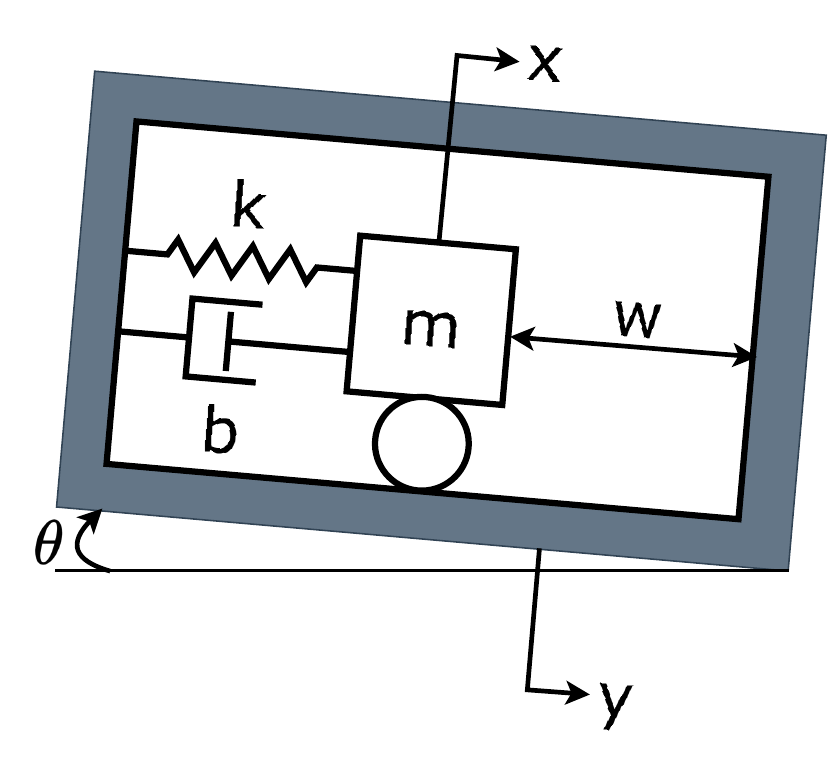}}
\subfloat[]{\includegraphics[scale=0.3]{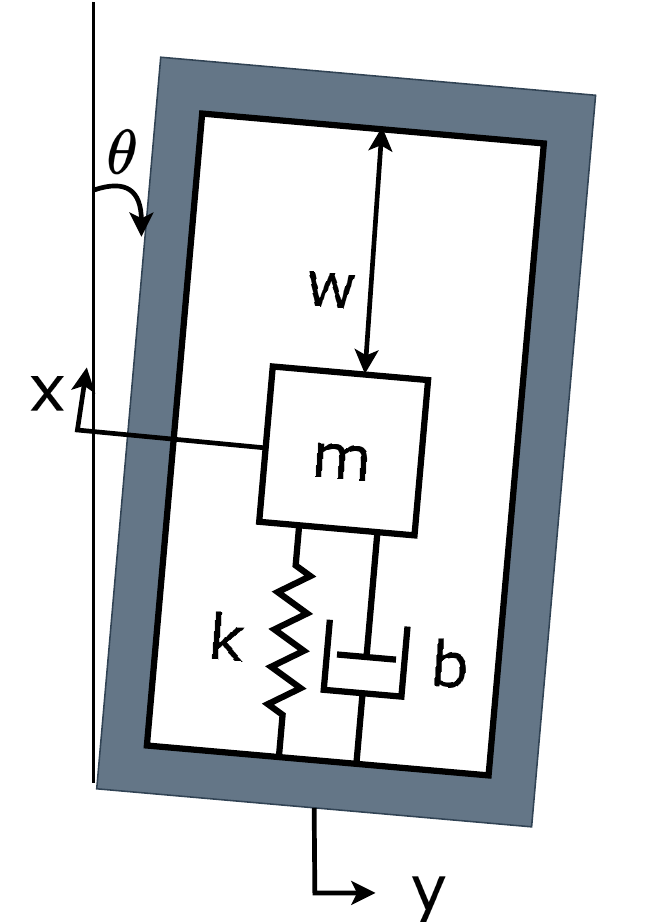}}
\caption{Schematic of a generic inertial sensor operating in the horizontal \textbf{(a)} and vertical \textbf{(b)} degrees of freedom, with induced inclination with respect to ground.}
\label{horiz}
\end{figure}

To better understand why tilt-to-horizontal coupling is a main source of noise for inertial sensors at lower frequency, we refer to Fig. \ref{horiz}(a) (see also \textit{Matichard and Evans, 2015} \cite{Matichard2015b}). 

An inertial sensor makes use of a mass and a soft spring inside a box to measure the relative displacement with respect to an inertial reference frame. Typically, in the presence of ground motion, we measure the movement of the box with respect to the mass. The softness of the spring defines the sensitivity of the sensor to specific frequencies. When rotation around the center of mass of the sensor occurs, we have an additional term appearing in its equations of motion, which for horizontal sensors we can write as:

\begin{equation}
    F_{\rm tilt} = m \cdot g \cdot \sin\theta
\end{equation}

where $m$ is the inertial mass, $g$ is the gravitational acceleration and $\theta$ is the angle of inclination. The equations of motion of a damped spring-mass system, including the term induced by the inclination become:

\begin{equation}
    F = m\ddot{x} = -b\dot{w} -kw + F_{\rm tilt}
\end{equation}

where $x$ is the direction of motion of the mass, $w$ is the difference in motion with respect to the reference frame ($w = x-y$), $k$ and $b$ are the coefficients of stiffness and damping of the spring, respectively. As stated above, the inclination of interest is very small, so we can assume $\sin \theta$ $\sim$ $\theta$ and study the system in the frequency domain, applying the Fourier transform:

\begin{equation}
    m(w+y)\omega^2 = -ib(w)\omega - k(w) + mg\theta
\end{equation}

\begin{equation}
    w(\omega^2m + i\omega b + k) = -\omega^2 y m + mg\theta
\end{equation}

Solving for $w$ we obtain:

\begin{equation}
    w = \frac{m\omega^2}{m\omega^2+i\omega b+k}(-y+\frac{g}{\omega^2}\theta).
    \label{solution2}
\end{equation}

\begin{figure}
    \centering
    \includegraphics[width=0.7\linewidth]{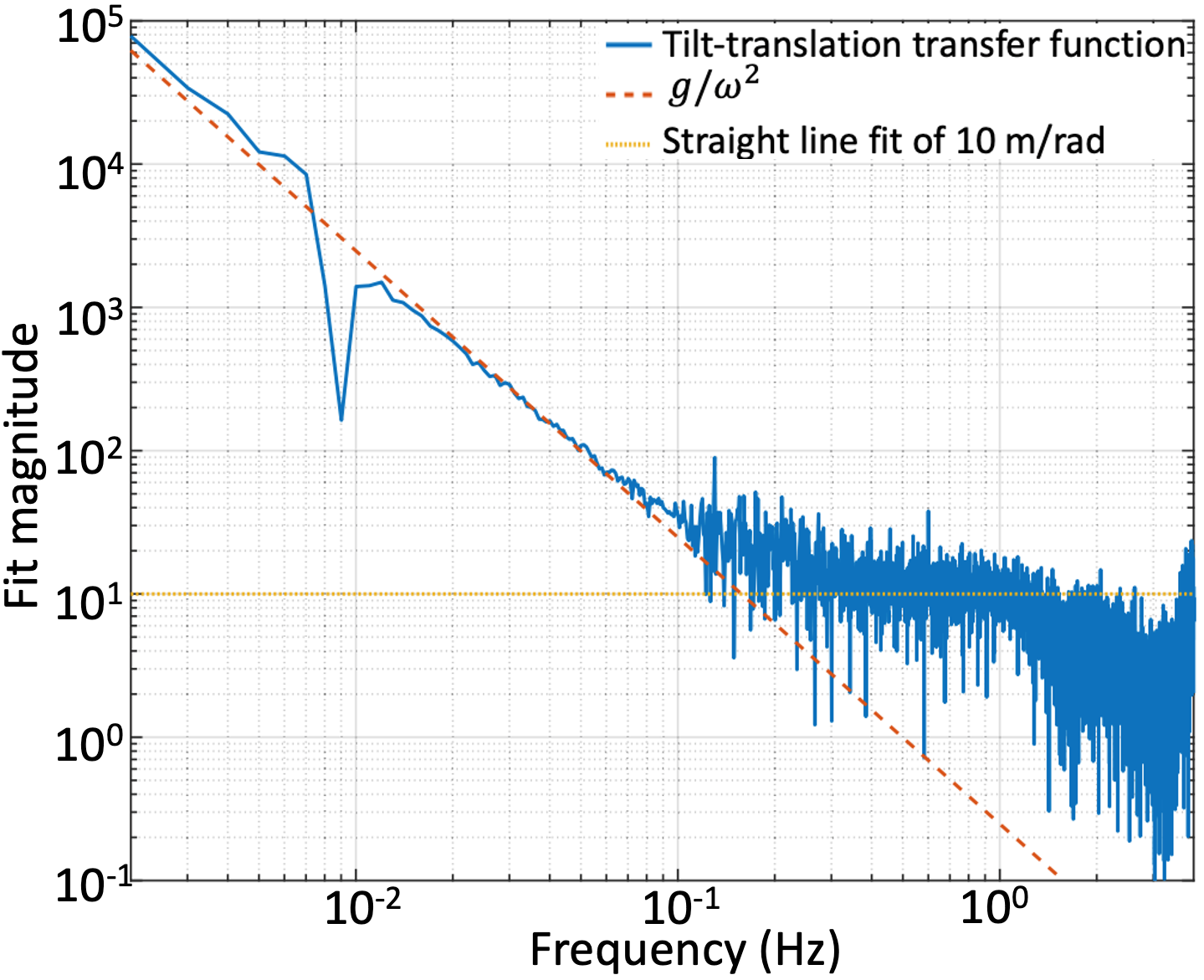}
    \caption{Tilt to translation transfer function example. This plot shows the relation between the translation measured by a seismometer and the tilt measured by the Beam Rotation Sensor developed by \cite{Venka2017a} in 2017. It has the expected g/$\omega^2$ relation at low frequencies and also shows a linear dependence at higher frequencies. Figure taken from \textit{Venkateswara et al., 2017}\cite{Venka2017a}.}
    \label{fig:ttl}
\end{figure}

The term in brackets includes the rotational effect, which is proportional to $g/\omega^2$, i.e. inversely proportional to the squared frequency. Fig. \ref{fig:ttl} shows an experimental example of this effect. This explains the dominance of the tilt effect at lower frequencies in horizontal inertial sensors, which is coupled and summed to the sensor's transfer function. Measuring the inclination and later subtracting its effect from the transfer function is the natural strategy to address this effect, however we will see that measuring very small angles ($<$10$^{-9}$ rad/$\sqrt{\rm Hz}$) is particularly challenging.\\

For inertial sensors operating in the vertical degree of freedom, the inclination is less impactful, and so it is its effect on the sensor sensitivity. If we consider the schematic in Fig. \ref{horiz}(b), the inclination contributes as cos$\theta$ to the equations of motions and for small angles $\cos\theta$ $\approx$ 1.\\

Other types of sensors are located on the platforms. However, they are affected by rotational motion in a different way. Briefly, on GW detectors, relative position sensors are of common use and include Capacitive Position Sensors (CPSs) \cite{Matichard2015} and Shadow Sensors such as the OSEM (Optical Sensor and ElectroMagnetic actuator) \cite{osem} and its upgrades \cite{Cooper2023, Carbone2012}. Despite position sensors being also affected by tilt, they are employed to measure the position of an object relative to a fixed local reference frame, rather than relying solely on the direction of gravity, so its effect is mitigated by the cancellation of commode modes from external factors, such as vibrations \cite{Matichard2015b}.

\subsection{Tilt-to-Length coupling and cavity locking}
\label{ttl}
\begin{figure}
    \centering
    \includegraphics[width=0.8\linewidth]{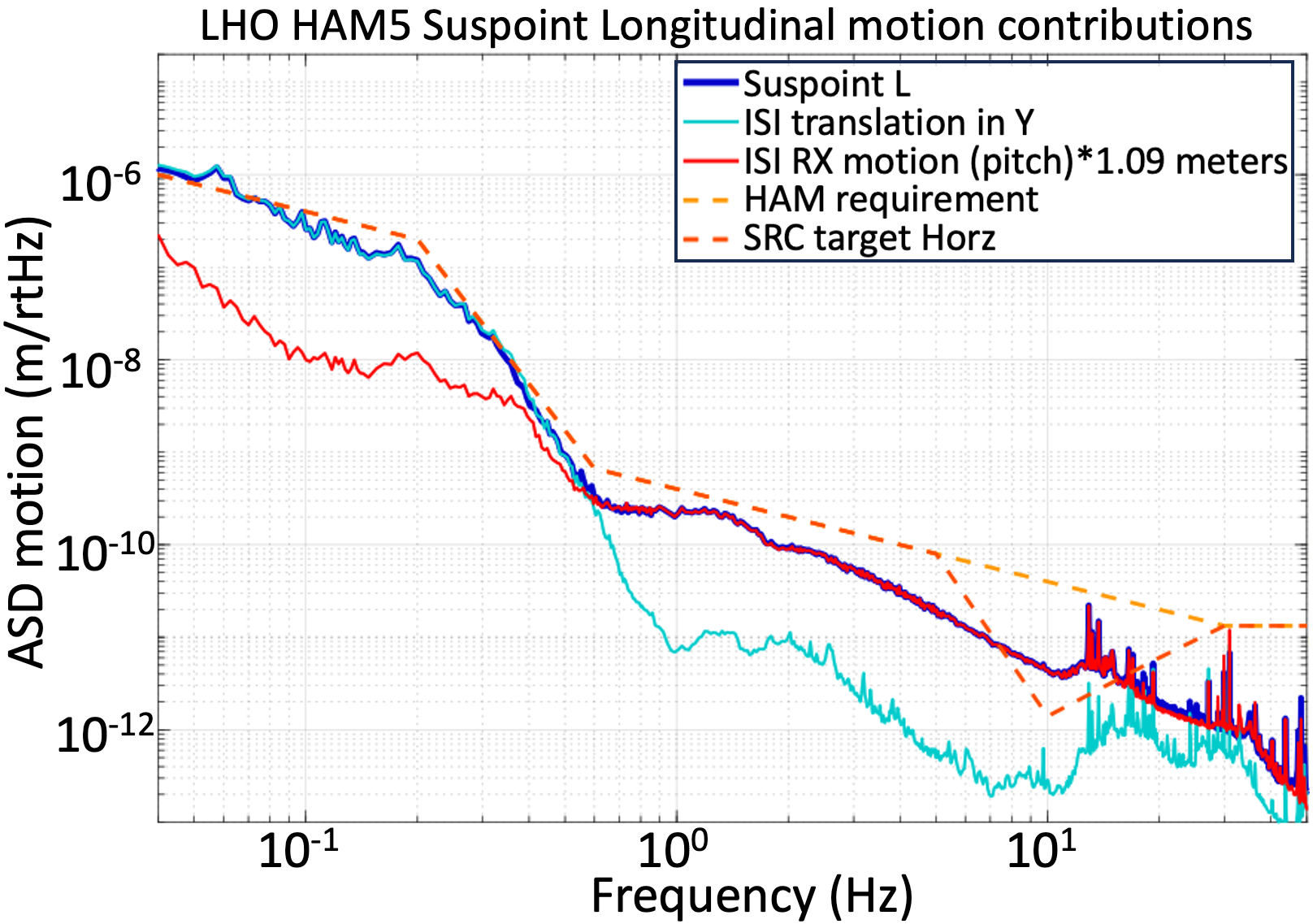}
    \caption{Contributions to the suspension point L motions. The platform of reference is the HAM5 at the LIGO Hanford site, on 3rd March 2018. This plots shows how the suspension point motions is a good measure for evaluating platform performance in X and Y motion. It show, also, the pitch (Rx) dominating motion above 1 Hz (red line). This line is a projection of the ISI pitch motion to suspension point, obtained by multiplying the data to the heigh of the suspension. In this example, the suspension analyzed is the Signal Recycling Cavity (SRC). Figure taken from \textit{Lantz et al., 2018} LIGO technical report\cite{dccT1800066}}.
    \label{fig:ligo_sus_pitch}
\end{figure}

Rotational ground motion affects the suspensions of the optics of aLIGO because the platform hosting them tilts, causing a displacement of the suspension point (i.e. where the optic's suspension is supported). Detectors like Virgo and KAGRA, on the other hand, rely almost entirely on passive isolation for their test masses (see Sec. \ref{IPs}). The suspension point motion is the best measure of optic motion and of platform performance and it was characterized on aLIGO in 2018. It was found that the rotational motion of the platform (typically pitch) tends to dominate at suspension points above 1 Hz (See Fig. \ref{fig:ligo_sus_pitch}). Addressing the motion here is important because it causes a variation of the distance between them, contributing to cavity length variations. The effect of rotational motion on the suspension points of test mass pendula or auxiliary optics lying on tables is the same (a variation of cavity length), as we illustrate in the following.\\
The longitudinal degree of freedom for the individual optics is defined as normal to the high reflective (HR) face of the optic. This depends on how it is aligned with respect to the reference axis. Thus, for example, an optic is aligned with positive X direction in the chosen coordinate system, but the other one is aligned along the  negative X direction. This means that moving an optic in the positive longitudinal direction makes the cavity shorter.

\subsubsection{Auxiliary optics suspensions}
\label{auxiliary}
\begin{figure}
    \centering
    \includegraphics[width=0.6\linewidth]{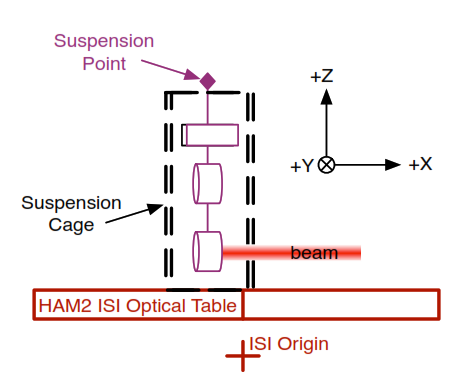}
    \includegraphics[width=0.6\linewidth]{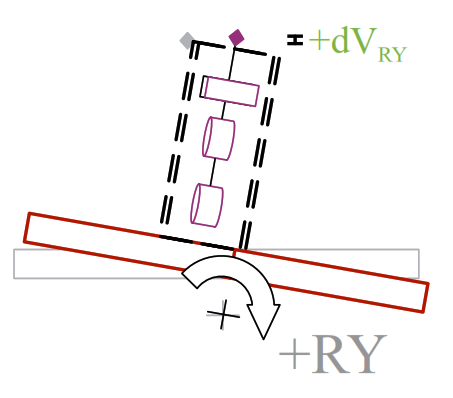}
    \caption{Side-view drawings example of the HAM-ISI hosting the Power Recycling Mirror (PRM) suspension, showing the resulting positive vertical displacement, +dV of the suspension point and platform tilt from looking along the +Y direction, with exaggerated +RY motion about the ISI coordinate center. The red rectangle is the outline of the ISI stage 1 optical table, and the red + is the center of the ISI cartesian coordinate system. The purple diamond is the suspension point for the PRM suspension chain (drawing not to scale). Figure adapted from \textit{Kissel et al., 2013} LIGO technical report \cite{dccT1100617}.}
    \label{fig:aux_cavity}
\end{figure}

Rotational motion on the platforms propagates and affects the suspensions of the auxiliary optics (i.e. the mirror contributing to power and signal recycling cavities, mode cleaning cavities, all suspended via multiple-staged pendula). On Virgo, the auxiliary optics lie on suspended benches (see Sec. \ref{IPs} later). The motion at the suspension points has been estimated on aLIGO in 2018, during Observation Run 2. Referring to the configuration in Fig. \ref{fig:aux_cavity}, a rotational motion along the Y axis of the platforms will induce a further motion along the X axis. Moreover, the effect will also result in a vertical displacement of the suspension point. If a suspended cavity lies on two platforms (each platform hosting a suspension), the effect will cause cavity misalignment and locking difficulties. For the cavity formed by two suspensions like the ones in Fig. \ref{fig:aux_cavity} on two different platforms, the difference in distance $\Delta L$ between the suspension points depends on the motion of the single platforms along the optics' direction, and the height of the suspensions:

\begin{equation}
    \Delta L = P1_x - P2_x + (P1_{Ry} - P2_{Ry}) \times H
\end{equation}

where $Pi_x$ is the motion of the platform along the X axis, $Pi_{Ry}$ the rotational motion along the Y axis and $H$ is the height of the suspensions.\\
On aLIGO, at this stage ground motion is important at the microseism, somewhat important from 1-10 Hz as the performance transitions from active to passive, and quite relevant above 10 Hz where most of the performance is passive. It was observed that on aLIGO this motion is dominated by pitch above 1 Hz.

\subsubsection{aLIGO test masses suspensions}
\label{ligo_susp}
\begin{figure}
    \centering
    \includegraphics[width=1\linewidth]{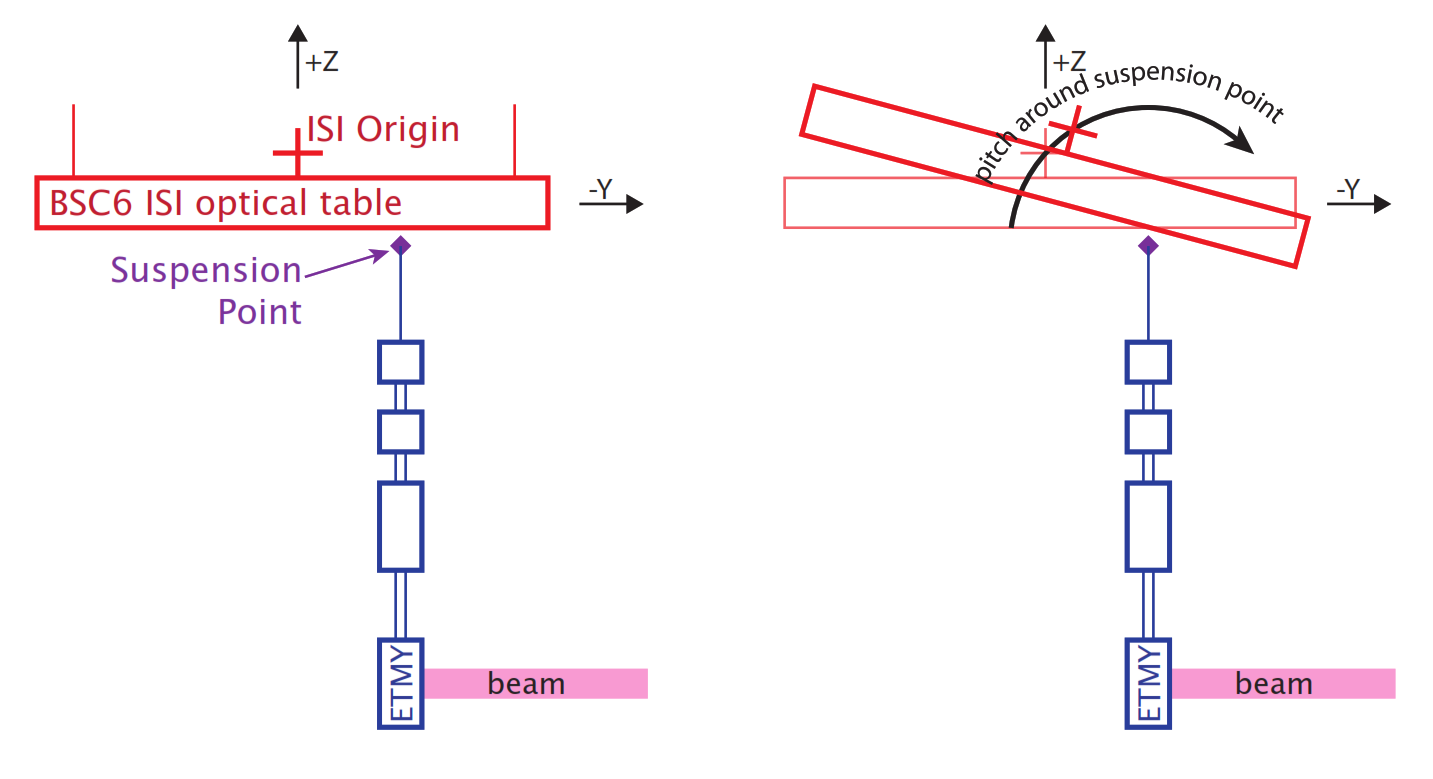}
    \caption{Side drawing of the BSC-ISI with End TM Y (ETMY), looking along the +X direction, showing exaggerated pitch motion about the ETMY suspension point. The red rectangle is the outline of the ISI stage 2 optical table, and the red + is the center of the ISI cartesian coordinate system. The purple diamond is the suspension point for the ETMY optical chain. If the optical table has positive pitch around the ETMY suspension point, the center of the ISI table will rotate in the +RX direction, and translate in +Z and -Y. Figure adapted from \textit{Kissel et al., 2013} LIGO technical report\cite{dccT1100617}.}
    \label{fig:ligotestsuspoint}
\end{figure}

aLIGO test masses are suspended from two stages of ISI platforms inside the BSCs, forming a quadrupole suspension which suspension point is on the upper ISI stage. We have already illustrated the effect of tilt on the sensors used to stabilize the ISI. In the case of the test mass, the suspension point motion is a reliable figure of merit for ISI motion. When tilt along the X axis happens, the center of the ISI rotates along the X axis and the suspension point translate in horizontal and vertical direction (see Fig. \ref{fig:ligotestsuspoint}). This created a difference in the cavity lengths, leading to cavity locking issues. 

Since the gravitational-wave signal is the result of the difference in arm lengths of the interferometer, and this is given by the difference in length of the two cavities forming the arms (DARM = (ETMX-ITMX) - (ETMY-ITMY)). Keeping the cavities stable from spurious length variations is then of critical importance.

%\begin{equation}
 %   DARM = (ETMX-ITMX) - (ETMY-ITMY)
%\end{equation}

\subsection{Tilt effect on inverted pendula}
\label{IPs}
\begin{figure}[h]                                             
    \centering
    \includegraphics[width=0.6\linewidth]{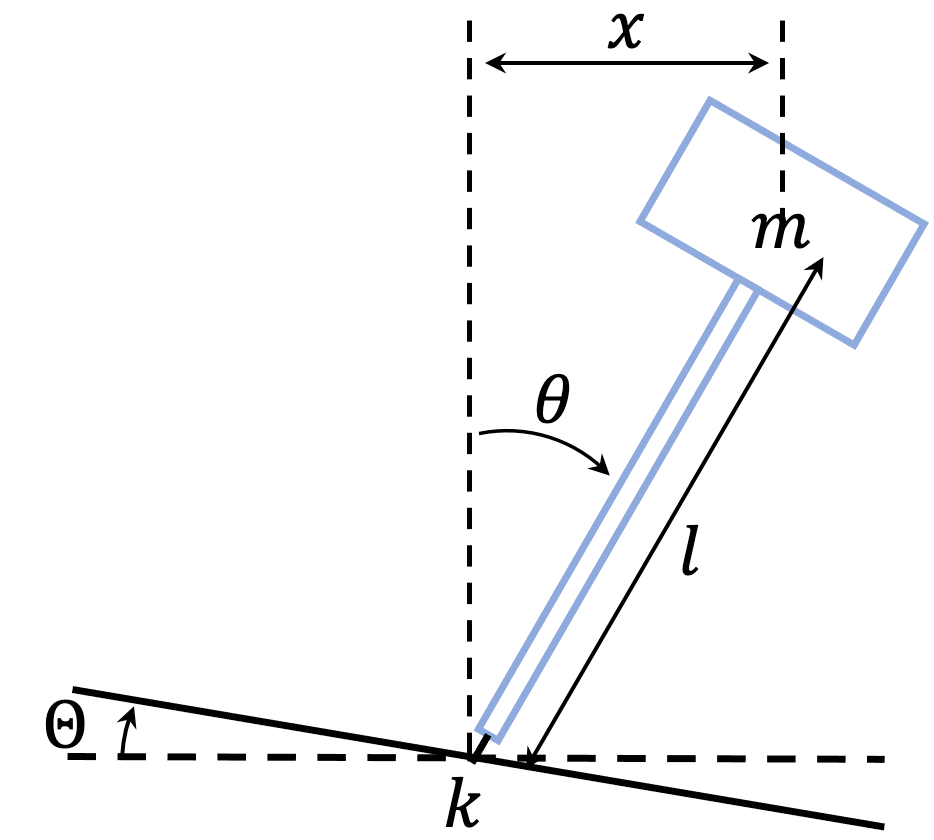}
    \caption{Schematic of an inverted pendulum subjected to ground tilt.}
    \label{fig:IP}
\end{figure}

Virgo and KAGRA detectors both utilize inverted pendula (IPs) to passively isolate the test masses. An inverted pendulum (the so called \textit{minitower} \cite{acernese2015advanced}) is also utilized on the suspended benches of Virgo to isolate the auxiliary optics. An inverted pendulum is a pendulum that has its center of mass above its pivot center, making it an unstable system. Inverted pendula can be maintained stable thanks to a suitable control system applied at the pivot point when they start to fall. The equations of motion of an inverted pendulum with a fixed pivot point are similar to that of an uninverted pendulum. Assuming absence of friction, pendulum length $l$, pivot angle $\theta$ and gravitational acceleration $g$, we obtain

\begin{equation}
    \ddot\theta - \frac{g}{l}\sin\theta = 0
\end{equation}

which states that that the pendulum is accelerating away from his pivot point, and longer pendula fall slower. When an inverted pendulum is subjected to tilt, this happens at the pivot point. Considering the inverted pendulum of mass $m$ and leg of length $l$ and negligible mass in Fig. \ref{fig:IP}, we can model a more realistic pivot point as a spring of stiffness $k$ and damping $b$ connecting the leg to the floor. The pivot tilt of the pendulum is still $\theta$, the floor tilt is $\Theta$. 

The above equation, becomes

\begin{equation}
    I \ddot\theta + b(\dot\theta - \dot\Theta) + k(\theta - \Theta) - mgl\sin\theta = 0
\end{equation}

where $I= ml^2$ is the moment of inertia. These equation is similar to the one we solved above for the mass-spring system of the inertial sensors. For small pivot angles $\theta << 1$ we can apply the Fourier transform and obtain

\begin{equation}
    ml^2\omega^2\theta + ib\omega(\theta - \Theta) + k(\theta - \Theta) - mgl\theta = 0
\end{equation}

Looking at Fig. \ref{fig:IP}, we can write $\theta$ in terms of $x$ as $\theta = x/l$, being $x$ the displacement of the mass from its vertical location. The equation then reads

\begin{equation}
    ml\omega^2x = - ib\omega(\frac{x}{l} - \Theta) - k(\frac{x}{l} - \Theta) + mgx
\end{equation}

Manipulating the equation such that we get the terms with $x$ to the left and the terms in $\Theta$ to the right:

%\begin{equation}
 %   ml\omega^2x  + \frac{ib\omega x}{l} + \frac{kx}{l}  - mgx= b\Theta + k \Theta
%\end{equation}

%\begin{equation}
 %   x(-ml\omega^2 + \frac{ib\omega}{l} + \frac{k}{l} -mg) = \Theta(ib\omega + k)
%\end{equation}

\begin{equation}
    x(-\omega^2 + i\omega\frac{b}{ml^2} + [\frac{k}{ml^2} -\frac{g}{l}]) = \Theta(i\omega\frac{b}{ml} + \frac{k}{ml})
\end{equation}

We can define the term in square brackets as $\omega_0^2 = \frac{k}{ml^2} -\frac{g}{l}$. Substituting, we obtain

\begin{equation}
    x\omega_0^2(-\frac{\omega^2}{\omega_0^2} + i\frac{\omega}{\omega_0^2}\frac{b}{ml^2} + 1) = \Theta(i\omega\frac{b}{ml} + \frac{k}{ml})
\end{equation}

If we recall the definition of quality factor of an oscillator as the ration between the natural frequency $\omega_0$ and damping factor $\beta$, such as $Q=\omega_0/\beta$, we can define $\beta = b/ml^2$ and substitute in the previous equation, obtaining

\begin{equation}
    x\omega_0^2(-\frac{\omega^2}{\omega_0^2} + i\frac{\omega}{\omega_0Q} + 1) = \Theta(i\omega\frac{b}{ml} + \frac{k}{ml})
\end{equation}

From this point we can show that the contribution of tilt at the base of the inverted pendulum leads again to a tilt-to-length effect

%\begin{equation}
 %   \frac{x}{\Theta} = \frac{i\omega\frac{b}{ml}+\frac{k}{ml}}{\omega_0^2(-%\frac{\omega^2}{\omega_0^2} + i\frac{\omega}{\omega_0Q} + 1)}
%\end{equation}

\begin{equation}
    \frac{x}{\Theta} = \frac{\frac{k}{ml}(i\omega\frac{b}{k}+1)}{\omega_0^2(-\frac{\omega^2}{\omega_0^2} + i\frac{\omega}{\omega_0Q} + 1)}
\end{equation}

In low frequency conditions ($\omega<<1)$ the terms in brackets goes to 1 and we obtain

\begin{equation}
    \frac{x}{\Theta} = \frac{\frac{k}{ml}}{\omega_0^2}
\end{equation}

To better understand what this effect means physically, we can recall the definition of $\omega_0^2$ and use it to rewrite the term $\frac{k}{ml}=g+\omega_0^2 \cdot l$. In this case we obtain

\begin{equation}
    \frac{x}{\Theta} = \frac{g}{\omega_0^2} + l
\end{equation}

This derivation describes the tilt effect on an inverted pendulum as a contribution inversely proportional to the square of the natural frequency of the pendulum, and depending on the length on the pendulum leg. This derivation assumes that the ground tilts at the base of the leg, which is the condition of the inverted pendula installed on gravitational wave detectors. In the following we describe how they are deployed on Virgo and KAGRA to suspend the test masses.\\

\begin{figure}[h]
    \centering
    \includegraphics[width=0.8\linewidth]{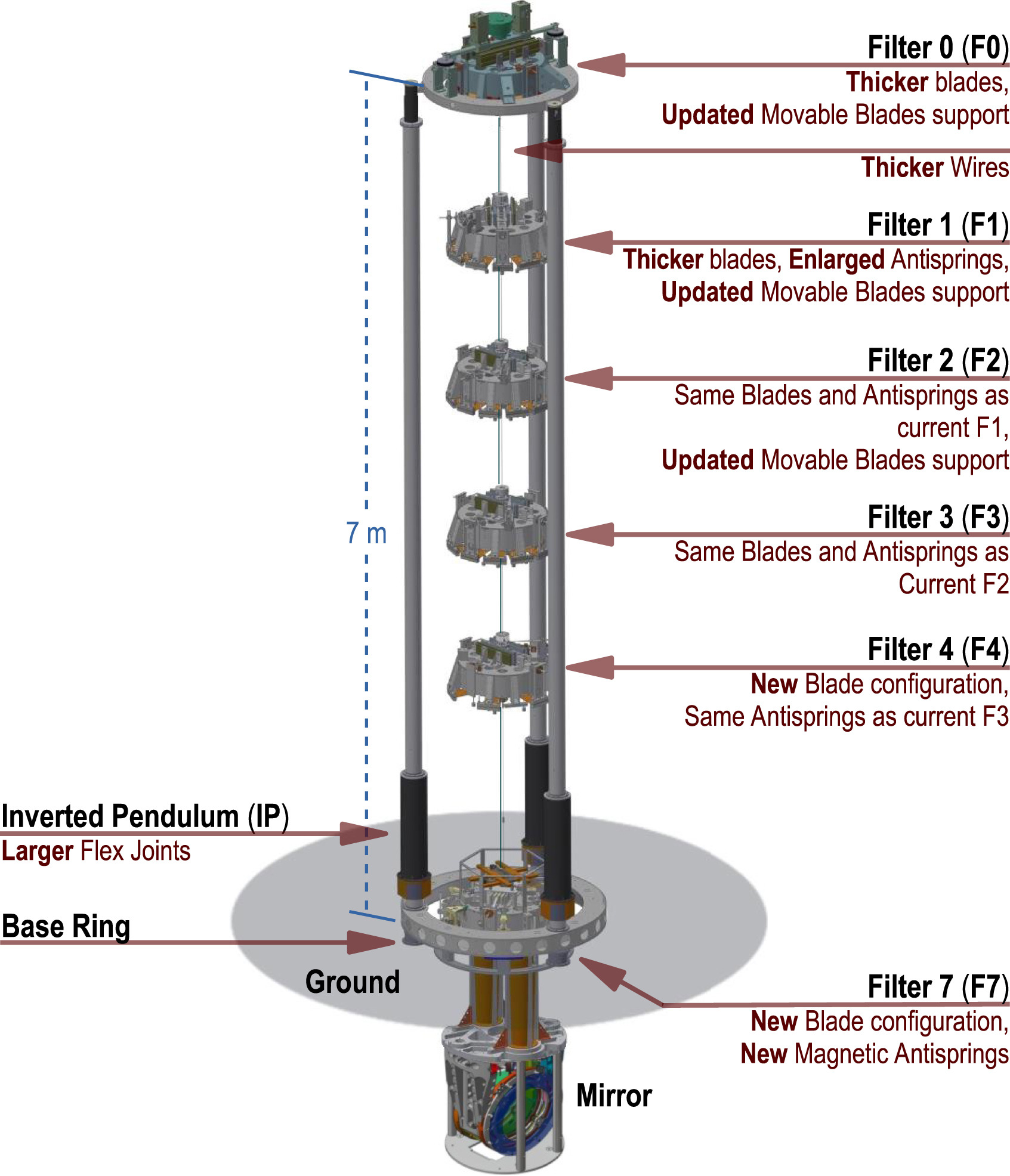}
    \caption{Image of the stages forming the Virgo Superattenuator (SA). Figure taken from \textit{Acernese at al., 2015}\cite{acernese2015advanced}.}
    \label{fig:VirgoSA}
\end{figure}

\paragraph{\textbf{Virgo detector}} Virgo is designed to observe in the 10 Hz to 10 KHz bandwidth and is provided with a passive isolation system involving a payload and a multistage pendulum called SuperAttenuator (SA) for each test mass and the beam splitter \cite{Dattilo2003, Acernese2010, Accadia2011}, shown in Fig. \ref{fig:VirgoSA}. The mirrors are kept suspended in a free-fall state in the interferometer plane. Each SA includes an inverted pendulum, a top mechanical filter, and a series of subsequent filters, leading to the payload at the bottom, which is equipped with a marionette and the actuation cage hosting the mirror. The payload is suspended directly from the last filter stage. Each filters consists of vertical oscillators made by blades and anti-springs. The details of each components of the SA are exhaustively explained in \textit{Accadia et al., 2011}\cite{Accadia2011}. Interestingly, it was found that on Virgo the main seismic noise contribution in the VIRGO detection band comes from the residual seismic vertical component coupled to the beam direction. This makes the contribution of the other degrees of freedom above 4 Hz negligible. In 2025 AdVirgo+ will enter a phase II, with major upgrades involving mirros mass increasing and a consequent revision of the SA \cite{Basti2023}.\\

\paragraph{\textbf{KAGRA detector}} KAGRA is designed to work underground to minimize sesimic and Newtonian noises \cite{Akutsu2018}. Its innovative configuration also makes use of cryogenic temperatures and sapphire test masses to reduce thermal noise \cite{Somiya2012, Michimura2017, Akutsu2019, Akutsu2021, Akutsu20212}. Therefore, its seismic isolation system must comply with low temperatures and underground environments. The test masses are suspended by a nine-stages vibration isolation system (named type-A suspension), of which the lower four stages are cooled at 20 K and are referred at as the cryogenic payload \cite{Ushiba2021}. Actuation is performed directly at the payload/mirror level \cite{Michimura2017}. In these detectors, a tilt of the upper filter induces a bend in the lower filter suspension wire that makes the lower filter translate. To evaluate the horizontal transfer function of the pendulum, the diagonal terms of the matrix of each stage should be multiplied. However, the tilt coupling makes the terms non diagonal and the multiplication method is used only for an estimation of the horizontal transfer function.\\

\begin{figure}[h]
    \centering
    \includegraphics[width=0.95\linewidth]{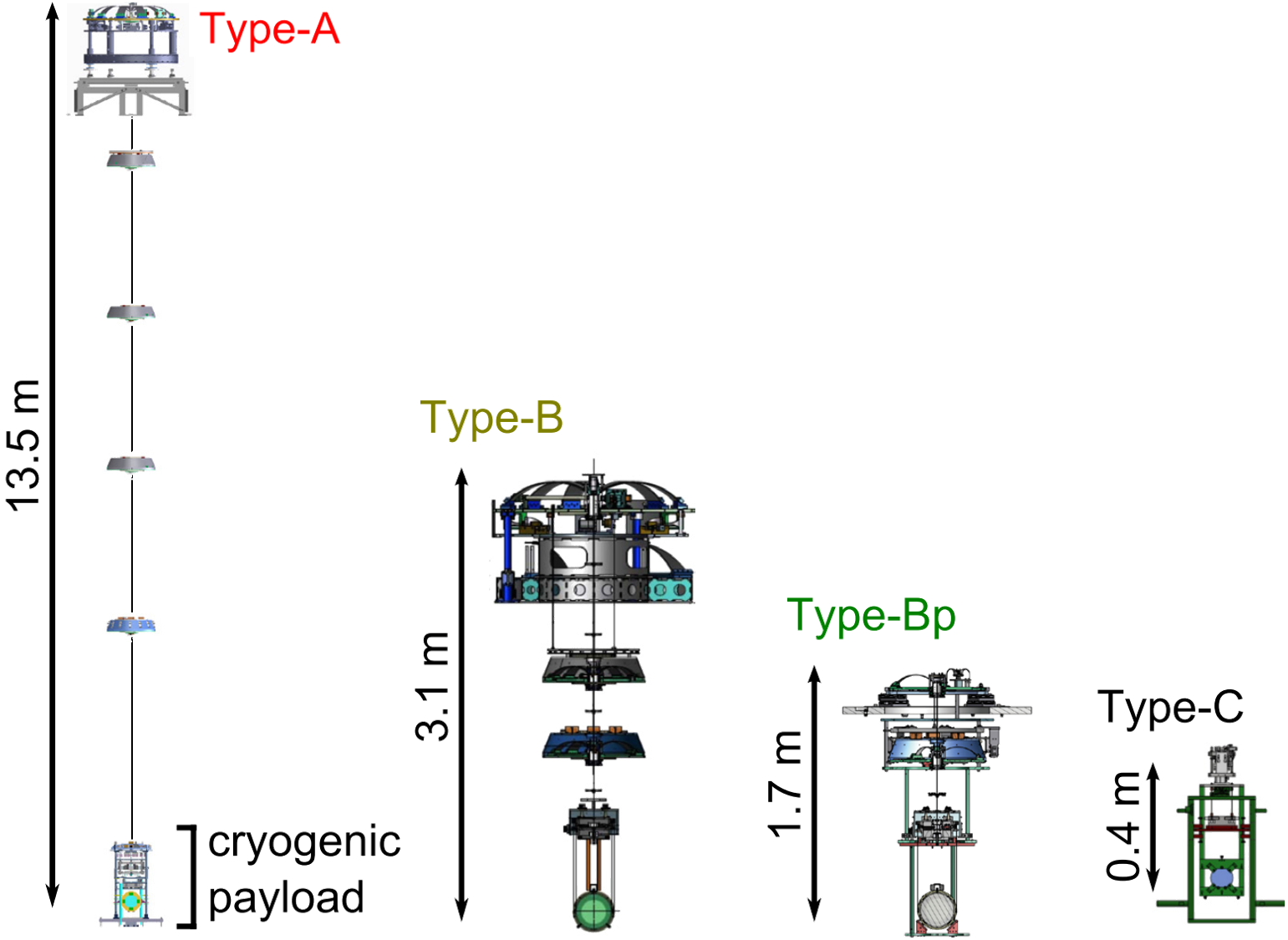}
    \caption{Four different types of KAGRA suspension system. Type-A is for test masses; Type-B is for beam splitter and signal recycling mirrors; Type-Bp is for power recycling mirrors and Type-C is for other auxiliary mirrors. Figures adapted from \textit{Akutsu et al., 2019}\cite{akutsu2019first}}
    \label{fig:kagrasus}
\end{figure}

In addition to the above, rotational motion can be induced in the test masses in other ways than seismic (control noise, radiation pressure, suspensions thermal noise, wind, etc.). Despite the different origin, the effect is the same - an unwanted movement of the test mass on the rotational degrees of freedom that impacts the sensitivity of the detector. The strategies to address these issues target both the source of the noise and the effect produced (the movement of the test mass). The former is beyond the scopes of this paper, while the latter concerns measurement strategies of angular displacements, which are relevant to this review.

\subsection{Detectors requirements}
\label{requirements}
\begin{figure}
    \centering
    \includegraphics[width=0.8\linewidth]{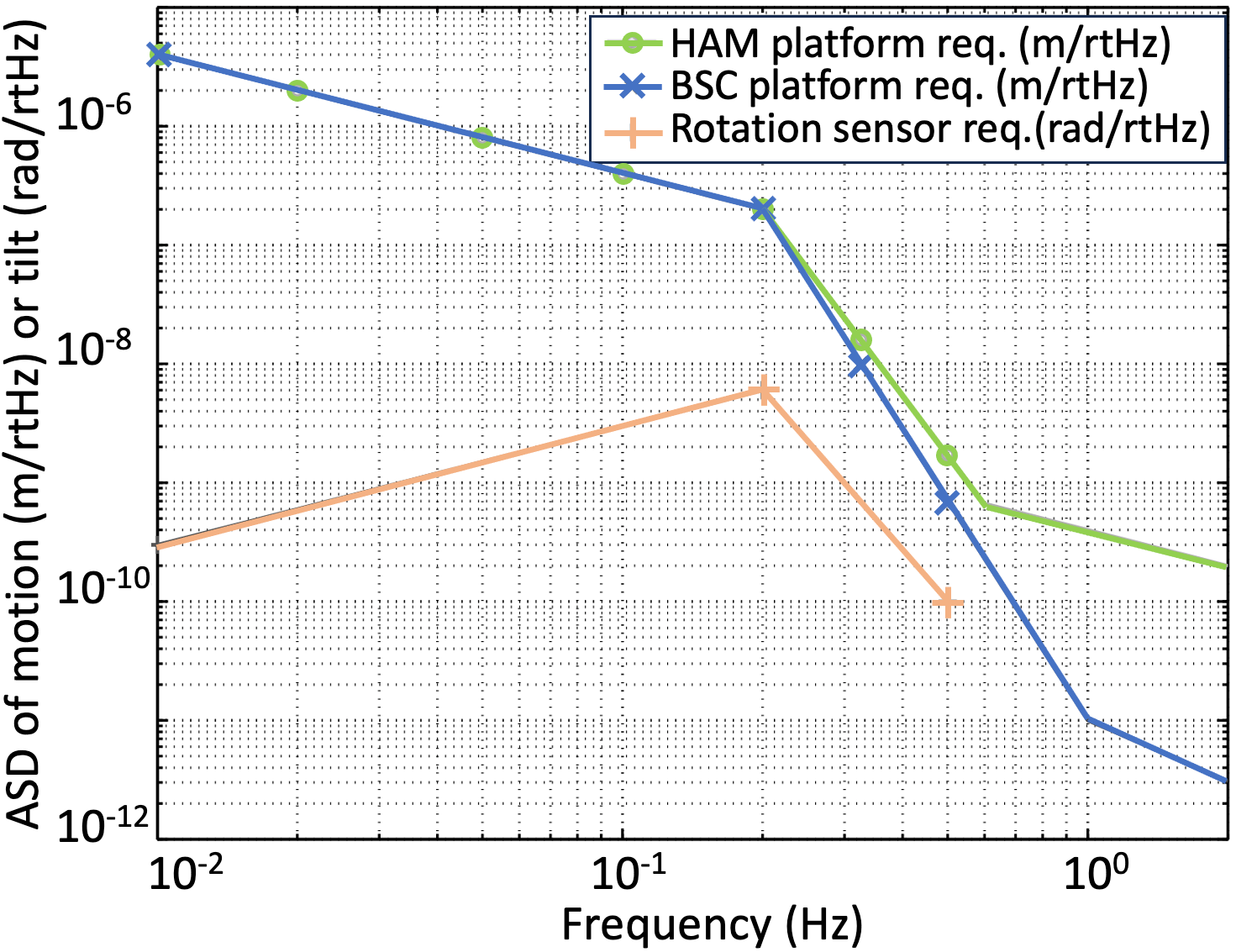}
    \caption{Performance requirement for the BSC and HAM isolation systems and noise requirement for aLIGO rotation sensors. Figure adapted from \textit{Lantz et al., 2009}\cite{lantz2009requirements}. }
    \label{fig:ligoreq}
\end{figure}

As we saw earlier, tilt affects different parts of the detectors, with consequences on the sensitivity at lower frequencies. For aLIGO, ideally the requirements include tilt sensing about the horizontal axes so that this signal can be removed in real time from the horizontal seismometers and only the true ground translation can be used to control the isolation system. \textit{Lantz et al., 2009} \cite{lantz2009requirements} provide details of the aLIGO requirements for rotation sensing, which we summarize in Fig. \ref{fig:ligoreq}. Virgo and KAGRA passive isolation systems allow the detectors to be more relaxed on the rotational degrees of freedom. We saw that on Virgo it is contributing below 4 Hz, while it can become important in bad weather conditions, due to wind-induced tilt of the suspensions. In this case, a tiltmeter installed at the top stage of the SA is required, with a sensitivity of 10$^{-8}$ rad/$\sqrt{\rm Hz}$ in the mHz band, because the accelerometer mounted there is unable to distinguish between real tilt motion and horizontal acceleration. KAGRA can benefit of the underground location to minimize the effect of ground motion, even induced by adverse weather conditions, on its suspensions.\\
From the controls point of view, the ultimate goal is to keep the cavities in resonance, i.e. maintaining low relative motion in the presence of differential ground tilt. In Sec. \ref{angular} we will see that there are basically three techniques to do this. One method is to filter the signals from the horizontal seismometers as a function of frequency and to use the signal in the band where the signal is dominated by true translation and reject the signal at frequencies where the tilt signals dominate. A second technique is to measure the angular acceleration directly. A third approach is to measure and directly control the distances between the support points of the optics (Angular Sensing and Control) \cite{robertson1982passive, aso2006active}. All these techniques require a robust sensing and control system, an adequate dynamic range and low control noise.

In order to account for angular displacements and satisfy the requirements of the detectors, substantial progress was made in the last decade to enhance the sensors for directly measuring tilt motion and to develop new strategies for controls.\\
In the following Sections, we will see the state of the art of these strategies and we will briefly illustrate the prospects for future generations GW detectors.

\section{Angular motion sensing and control}
\label{angular}
\textit{Matichard et al. 2015} \cite{Matichard2015} illustrated the technologies developed for GW detectors up to 2015. Substantial progress has been made in rotational seismology in the last decade for GW detectors improvement. In the following, we will describe the latest technologies developed for rotational motion sensing on GW detectors.\\
\textit{Matichard and Evans, 2015} \cite{Matichard2015b} presented an exhaustive instrument topology illustrating the different types of sensors used to estimate and subtract tilt. In recent years, the development of very low noise rotation sensors specifically dedicated to measure angular motion (and then apply feedback) has been preferred to indirect estimation and subtraction strategies. Although there is ample availability of rotational sensing devices that could potentially satisfy the sensitivity requirements for GW detectors, several constraints prevented their use. The most common is the need for compactness and ultra-high vacuum (UHV) compatibility, because the sensors are often located on the platforms in vacuum, where there is limited room and a tight weight budget (see for example the work by \textit{Sunderland et al., 2013} \cite{sunderland}, \textit{Jaroszewicz et al., 2018} \cite{Jaroszewicz2018}, \textit{Di Virgilio, 2020} \cite{di2020sagnac}, \textit{Bernauer et al., 2021} \cite{bernauer2021} and \textit{Kurzych et al., 2023} \cite{Kurzych2023}).\\

In general, the study of devices that can measure in all 6DOFs has been at the forefront in recent years. Several facilities have implemented vibration isolation systems to test technologies to be implemented on GW detectors. It is worth describing here some of these facilities, because at these sites, the most advanced technologies, such as sensors and actuators, as well as software systems, are developed and tested before installation onto the main detectors.\\

\paragraph{AEI 10-m Prototype}
The 10-m Prototype located at the Albert Einstein Institute (AEI) in Hannover (Germany) is a scaled-down facility for interferometric experiments \cite{Goler2010}. As per its original design, it is an ultra-low displacement noise facility consisting of an L-shaped ultra-high vacuum system with about 10 m long arms \cite{Dahl2012}. To achieve the desired seismic isolation, it is equipped with a seismic attenuation system (SAS) \cite{Dahl2012, Wanner2012, Bergmann2017, Bergmann2018} which then is planned to be paired to a suspension platform interferometer (SPI) \cite{Dahl2010, Dahl2012b, Sina2018}, to account for the differential drifts in position of the optical tables at frequencies below the fundamental resonance frequency of the SAS, that are due to mechanical tolerances from the manufacturing and assembling processes. While the SAS system is an active platform designed to isolate the optics on top of it from vertical and horizontal motion and each SAS platform is independent from the other \cite{Kirchhoff2020}, the SPI measures and control via feedback loop the differential motion between the three SASs, making them move as a single platform.\\

%\paragraph{MIT prototype}
\paragraph{Australian High Optical Power Facility}
The High Optical Power Facility (HOPF) located in Western Australia is an 80-m-long interferometric setup and is equipped with a vibration isolation system for technology tests. It was initially designed to test high-power technologies for GW detectors \cite{Ju2004, Zhao2006} and over the last two decades it developed a high-performance compact vibration isolation system, in order to comply with modern research on suspensions and seismic noise in GW detectors \cite{Chin2006, Barriga2009}. It consists in a triple pendulum on top of which four inverse pendulum legs supporting a square table provide a first stage of preisolation in horizontal translation. A second stage of preisolation is provided by a spring system isolating in the vertical direction. The whole preisolation stage has a resonance frequency of 100 mHz, slightly tunable \cite{Barriga2009}. Local sensing and actuation is applied on the preisolation stage. At the bottom of the pendulum, the test mass is suspended from a control mass which is equipped with sensing and actuation devices for mirror control in cavity locking experiments \cite{Barriga2009}.\\

%\paragraph{GEO600}
%NOT SURE TO KEEP THIS PART\\
%GEO600 is an interferometric gravitational-wave detector built by a German-British collaboration. The project pioneered the field of study of technologies for the following generations of detectors and it has operated at higher frequencies ($>$ 1KHz) \cite{Hild2006, Dooley2016}. Therefore, the vibration isolation system is designed to reduce seismic noise up to about 100 Hz, i.e. at the limit of the designed detector observational bandwidth \cite{Lck1997, Willke2002}. It is equipped with triple suspensions with a resonance frequency below 10 Hz, which isolate the test masses against seismic motion. \textit{Willke et at., 2002} \cite{Willke2002} describes in detail the design of the system. GEO600 experienced several upgrades over the last two decades, to enhance the performance of the detector at high frequencies. Therefore, the upgrades to the vibration isolation system mainly concerned the control system for local optics control \cite{Grote2010, Lck2010} and enhancements to reduce thermal noise \cite{Affeldt2014}.

\subsection{Technology state-of-the-art}
\label{sensing}
In this Section we report the recent technological progress developed by the facilities described above and by specialized laboratories. We will summarize the strengths and the limitations of the technologies, relatively to the requirements for GW detectors or GW testbed facilities.

\subsubsection{Technology based on inertial sensors}
\paragraph{\textbf{6DOF sensors}}
Inertial sensors are commonly used on GW detectors, thanks to their relatively small size, UHV compatibility and ease of use. They are generally suitable for modifications and hybridization with optical readouts to enhance their sensitivities \cite{Cooper2018}. As we saw in Sec. \ref{tilt}, horizontal inertial sensors are highly affected by inclination. The recent studies then focused on how to utilize combined inertial sensing to create a 6DOF sensor. These technologies use interferometric readouts to monitor the motion of a suspended mass relatively to the supportive platform, in all degrees of freedom.\\ 
The research conducted at the University of Birmingham and Nickef \cite{Ubhi2022, Prokhorov2024, DiFronzo2024} is an example of a 6DOF sensor for installation on GW detectors (Compact-6D \cite{ubhi2022six, Smetana2025}, OmniSens \cite{omnisens}). These devices make use of a mass suspended by a single fused silica fiber and low noise interferometric readout. 6D device can measure 10$^{-9}$ rad/$\sqrt{\rm Hz}$ at  0.1 Hz, which outperforms the current aLIGO inertial sensors in measuring angular degrees of freedom \cite{Prokhorov2024}.\\

\begin{figure}[h]
    \centering
    \includegraphics[width=0.6\linewidth]{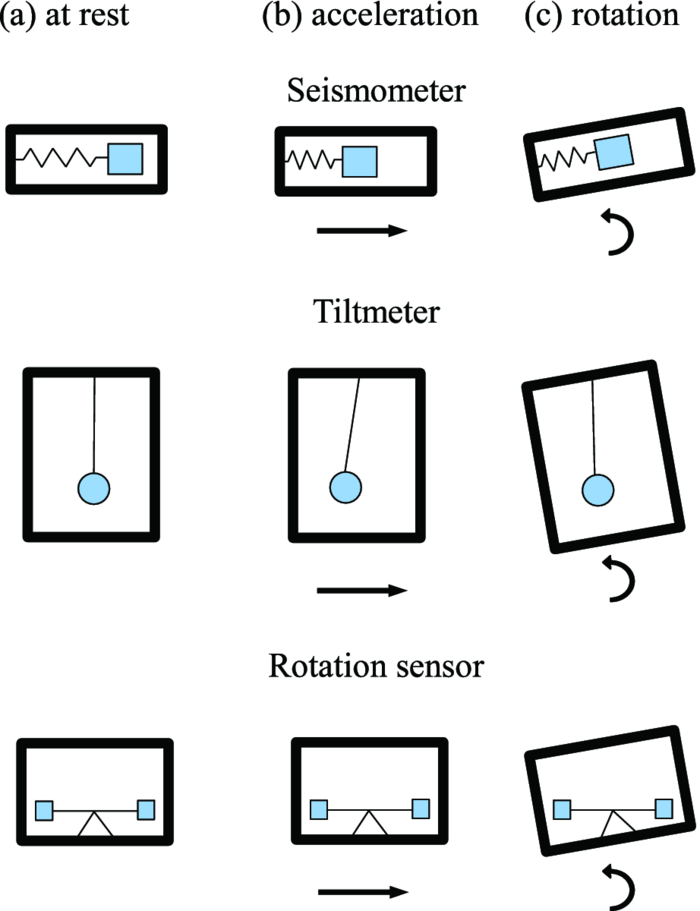}
    \caption{Schematic representation of the behavior of rotation sensors when tilt occurs, compared to what happens to the mass of an inertial sensors in the same condition. Figure taken from \textit{Venkateswara et al., 2014}\cite{Venkateswara2014}.}
    \label{fig:rotations}
\end{figure}

\paragraph{\textbf{Rotation sensors}}
Rotation sensors are also based on the concept of mass motion in a reference frame similar to the seismometers, but where the mass being instead, either a vertical pendulum (tiltmeters) or a balanced beam. For these type of sensors, tilt displacement is referred to the angular deflection of a simple pendulum with respect to its suspension platform or enclosure \cite{Venkateswara2014}. These devices suffer from the same problem that inertial sensors have, because they address tilt by comparing horizontal acceleration with the vertical. Fig. \ref{fig:rotations} is a schematic example of how these devices behave compared to translational inertial sensors, in the presence of tilt. The fundamental difference is that in rotation sensor the inertial mass is used to measure angular velocity, i.e. how much the inertial mass is tilting.

\begin{figure}[h]
    \centering
    \includegraphics[width=0.65\linewidth]{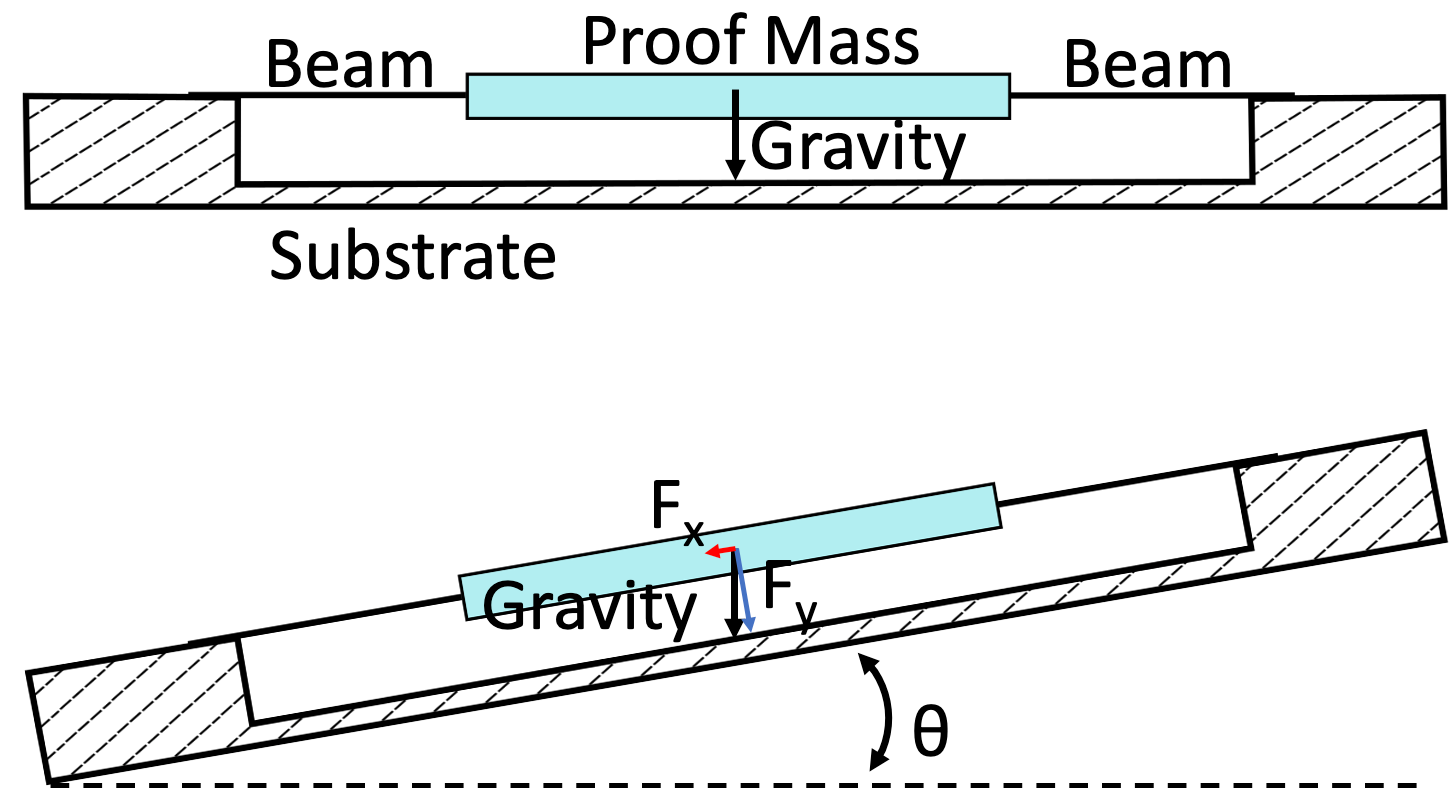}
    \caption{BRS working principle. High-precision devices of this type makes use of a heavy proof mass and a soft pivot (like a crosse flexure). When there is a rotation, inertia makes the proof mass move or experience torque. In most recent devices, the angular difference between the proof mass and the platform is measured by optical sensors.}
    \label{fig:brs}
\end{figure}

In order to overcome these limitations and exploit the local gravitational vertical frame of the mass, in the last decade the research by \textit{Venkateswara et al., 2017} \cite{Venka2017a} led to an improvement of Beam Rotation Sensors (BRS) when paired to extra devices. In BRSs, ground tilt is defined as its angle with respect to the horizontal axis of a nearly inertial frame fixed to the local gravitational vertical (defined by a free‐falling mass). This frame is not completely inertial because of variations of local gravity and Earth's rotation, but at low frequency (above 10 mHz) this effect is negligible. Fig. \ref{fig:brs} shows the basic principles of a BRS.

The BRS currently installed on aLIGO is paired to a seismometer and an autocollimator readout to subtract tilt and provide a "pure" displacement measurement. This sensor combination has an intrinsic sensitivity of 0.15 $\times 10^{-9}$ rad/$\sqrt{\rm Hz}$ above 100 mHz \cite{Venka2017a, Venka2017b}.\\
Since then the research on improving BRSs includes the device ALFRA developed in 2021 by \textit{McCann at al., 2021} \cite{McCann2021}. In ALFRA, The angle difference between the ground and beam is measured using a laser walk-off sensor (described in the next paragraph). ALFRA has a sensitivity of 0.1 $\times 10^{-9}$ rad/$\sqrt{\rm Hz}$ above 20 mHz. Upgrades of ALFRA are currently being tested to be installed on HOPF to enhance the facility vibration isolation system in the next 5 years \cite{Satari2022}.\\

\begin{figure}[h]
    \centering
    \includegraphics[width=0.6\linewidth]{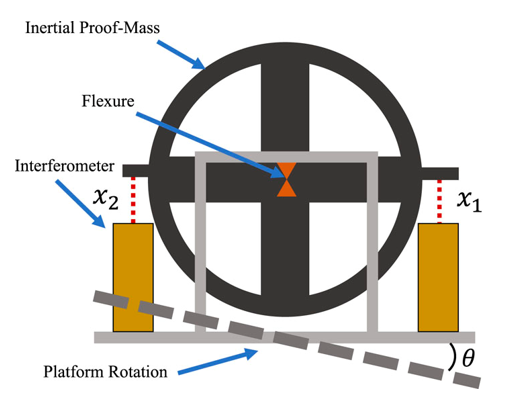}
    \caption{Schematic design of the CRS, as a general example of these type of sensors. Figure taken from \textit{Ross et al., 2023}\cite{ross2023}.}
    \label{fig:crs}
\end{figure}

Another family of devices based on inertial sensing is compact accelerometers. Generally, an accelerometer measures the rate at which the velocity of the proof mass changes - i.e. when it experiences a movement or gravity. In case of rotation accelerometers, the quantity measured is the change of velocity when the mass tilts. Tilt angles are then computed relative to ground. For GW detectors, this type of sensors have been optimized and adapted in size to fit resolution and installation requirements.\\
Among them there is the Cylindrical Rotation Sensor (CRS) developed by \textit{Ross et al., 2023} \cite{ross2023}, which aims to update the current Beam Rotation Sensors in use on aLIGO \cite{Rossth}. Fig. \ref{fig:crs} shows its schematic features.\\

This device is a 30-cm scale inertial rotation sensor that reaches a sensitivity of $\sim$ 5$\times$ 10$^{-12}$ rad/$\sqrt{\rm Hz}$ at 1.5 Hz;\\
on a similar design, the Quartz Rotation Sensor (QRS) developed by \textit{Venkateswara et al. 2021} \cite{venka-quar} is an accelerometer designed on a similar concept of the CRS that senses rotational torque with an inherently digital, load‐sensitive resonant quartz crystal. The QRS can measure $\sim$ 45$\times$ 10$^{-11}$ rad/$\sqrt{\rm Hz}$ around 1 Hz.\\

\begin{figure}[h]
    \centering
    \includegraphics[width=0.7\linewidth]{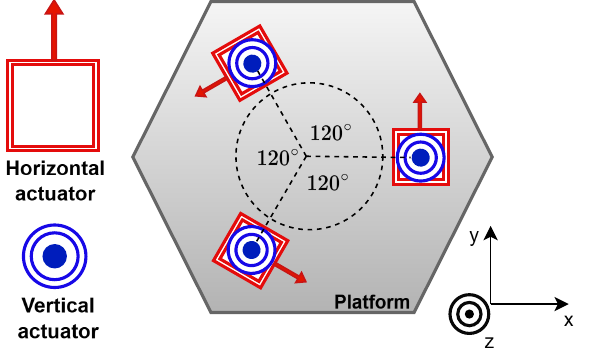}
    \caption{Design principle of the SILENT platform actively isolated in all 6 degrees of freedom through the scheme shown in the sketch. Figure adapted from \textit{Lakkis et al., 2025}\cite{Lakkis2025}.}
    \label{fig:platforms}
\end{figure}

\paragraph{\textbf{Active platforms}}
Other technologies involve the combination of multiple inertial sensors in various degrees of freedom to form a 6 DOFs isolation platform (E-TEST, SILENT \cite{etest, silent}).\\
While the working principles of these two concepts (single device and isolation platform) are both based on inertial sensing, the former prioritizes the compactness and ease of installation of a single device and precision angular motion measurements, while the latter relies on highly performative inertial sensors in vertical and horizontal DOFs for high precision motion measurement for stabilizing an isolation platform, without specifically targeting tilt motion but generally improving isolation from ground motion. Fig. \ref{fig:platforms} shows the basic principle of this concept using as an example the design of the SILENT platform. The rotational degrees of freedom of SILENT can be measured up to $\sim$ 2$\times$ 10$^{-9}$ rad/$\sqrt{\rm Hz}$ around 1 Hz \cite{Lakkis2025} and we will see later the motion suppression performance.

While the single device is ideal for installation on any GW detector, the latter technology is mainly developed for active isolation on future generations GW detectors, since it would require major changes for installation on current detectors. Both technologies aim for high-precision motion measurement for feedback control.\\

Single devices are designed to be installed on GW detectors where needed, thanks to their compact size \cite{crsdcc}. Initial tests on CRS at aLIGO in 2022 have shown improvement in tilt measurements compared with currently used sensors \cite{crsdcc2}. On Virgo, adverse weather conditions affect about 10\% of the Virgo duty cycle \cite{trozzophd} due to tilt motion. Hence, the CRS is also currently under study for possible installation on the inverted pendula of Virgo superattenuators (as per Virgo requirements and correspondence with Virgo members, the adaptation mainly concerns a reduction in size).\\

\begin{figure}[h]
    \centering
    \includegraphics[width=0.8\linewidth]{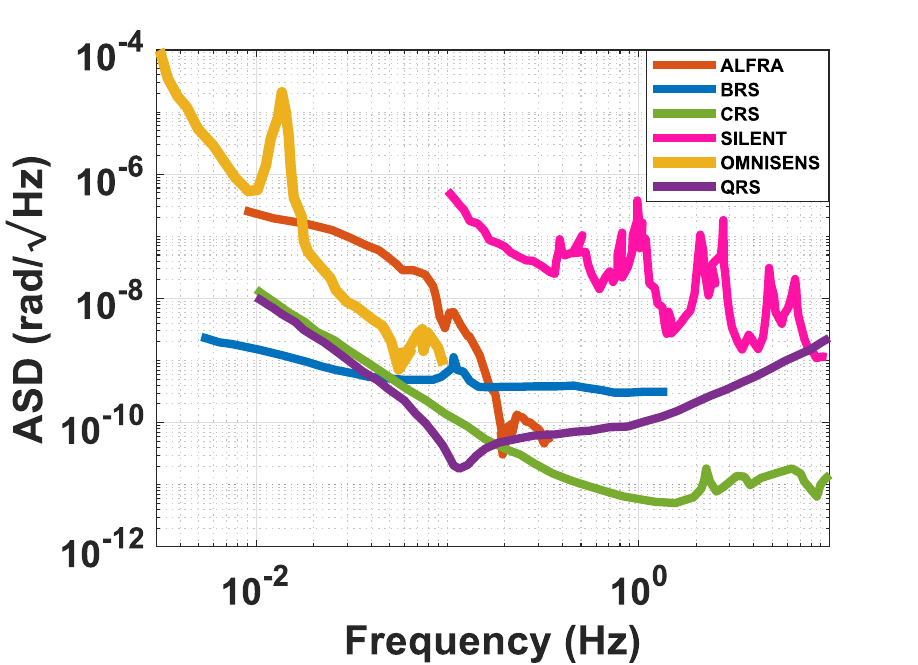}
    \caption{Summary of the noise floors of the technologies based on inertial sensors currently existing and tested.}
    \label{fig:inertial}
\end{figure}

Fig. \ref{fig:inertial} compares the performance of all the technologies described above with each other.

\subsubsection{Technology based on relative sensors}
\begin{figure}[h]
    \centering
    \includegraphics[width=0.6\linewidth]{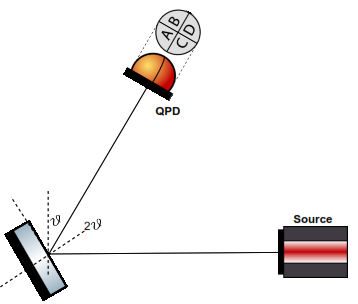}
    \caption{Basic principle of an optical lever: an optic is illuminated by a laser source and reflecting the light to a detector. When the optic is tilted by an angle $\theta$, the displacement is detected by the photodiode, typically a Quadrant Photodiode (QPD). The longer the lever arm, the smaller the measurable angle.}
    \label{fig:oplev}
\end{figure}

Relative displacement sensors are often used on GW detectors to measure displacements between objects (for instance, between two platform stages on aLIGO). Although they are not strictly speaking seismic sensors, directly measuring ground motion, they are worth mentioning because they measure the effect of ground tilt onto test masses and platforms on GW detectors.\\
Among the most common relative sensors, optical levers are well-known devices that are employed for non-contact measurements of angular displacements on GW detectors, relative to a reference point. Optical levers make use of a beam light and a position sensor to measure a small displacement and thus to make possible an accurate measurement of angles. Fig. \ref{fig:oplev} shows the working principle of a basic optical lever.

\begin{figure}[h]
    \centering
    \includegraphics[width=0.8\linewidth]{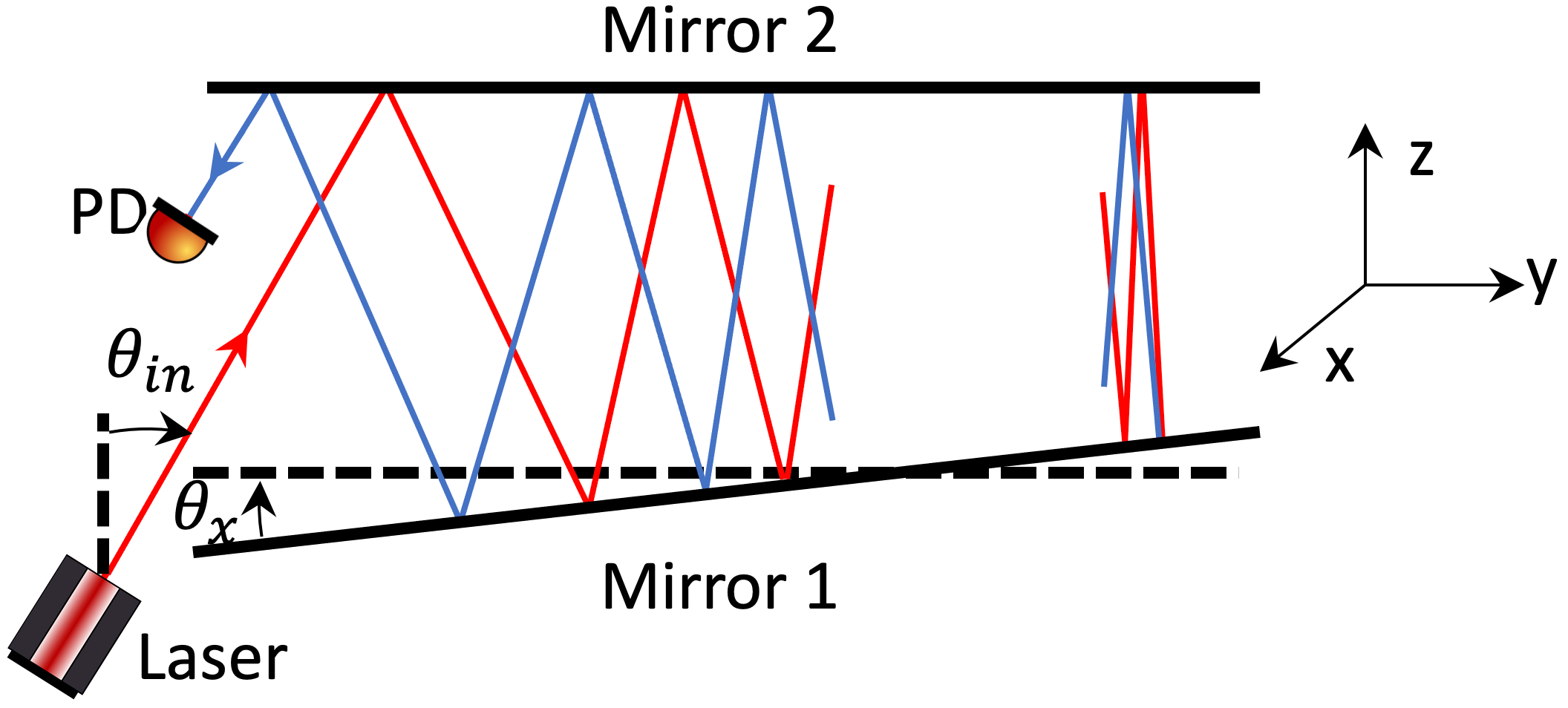}
    \caption{Schematic design and working principle of a multiple reflection lever. This has a similar working principle of a classic optical lever, overcoming the challenge of an extremely long lever arm by making the laser beam traveling long distance through a number of bounces between two mirrors proportional to the lateral displacement of the beam, increasing the sensitivity to the angle displacement. In the Figure, the laser source is injected at a know angle $\theta_{in}$. When the interested object (Mirror 1 in the Figure) tilts of an angle $\theta_x$, the beam spot shifts across the active area of a photodetector (PD). Intermediate reflections omitted for clarity.}
    \label{fig:wos}
\end{figure}

While in the past decades the research focused on improving the sensitivity of these devices, in the recent years the technology was adapted to be paired to sensors to improve their sensitivity, as in the case of the Walk Off Sensor (WOS) developed in 2001 by \textit{Zhou et al., 2001} \cite{Zhou2001} and perfected to be used on ALFRA \cite{McCann2021} as a readout device in 2021. This device shares a similar working principle with an optical lever, but makes use of multiple beam reflections to increase the lever arm. Fig. \ref{fig:wos} shows the schematic design of this technique. The WOS noise floor reads up to about 10$^{-9}$ rad/$\sqrt{\rm Hz}$ at 1 Hz.

More recently, technologies based on optical levers are investigated to be added to existing sensing and control setups of GW detectors. For example, at the AEI in Germany, optical levers have been adapted to contribute to measuring and controlling the motion between the seismic platforms of the 10-m Prototype. The study by \textit{K$\ddot{o}$hlenbeck, 2018} \cite{Sina2018} led to the concept of the Suspension Platform Interferometer (SPI) \cite{Koehlenbeck2023}, where the differential motion between platforms is targeted and addressed. The SPI makes use of interferometers measuring length displacements. Optical levers are included to account for angular displacements.
%Fig. \ref{fig:spi} shows the working principle of this setup, with a focus on the optical lever contribution.
The SPI angular sensitivity is about 10$^{-9}$ rad/$\sqrt{\rm Hz}$ at 1 Hz and the error signal can be suppressed to 20 10$^{-12}$ rad/$\sqrt{\rm Hz}$ at 1 Hz \cite{Sina2018}. The SPI concept (SPI Pathfinder) is currently under study for installation on aLIGO \cite{GWANW2025}, and an SPI setup is under development at the HOPF.

In 2018, \textit{Kokeyama et al., 2018} \cite{Kokeyama2018} developed a new type of angular sensor that was tested on KAGRA, adapted from a tabletop experiment realized in 2016 by \textit{Park et al., 2016} \cite{park2016}. This sensor is based on a Mach–Zehnder interferometer folded by a target mirror, with in- and quadrature-phase readout. It is called Interferometric Tilt Sensor (ITS).
The sensor fits in the non-contact, relative devices category, and has a RMS of less than 0.1 $\times$ 10$^{-9}$ rad/$\sqrt{\rm Hz}$, which is within the requirements for KAGRA mirror motion bandwidth. The large dynamic range and self-calibration features make ITS a competitive sensor to the classic optical levers.

%\begin{figure}[h]
 %   \centering
  %  \includegraphics[width=0.7\linewidth]{its.jpg}
   % \caption{ITS schematics. FIGURE TO BE SIMPLIFIED. ADD DETAILS IN CAPTION. Figure adapted from \cite{Kokeyama2018}.}
    %\label{fig:its}
%\end{figure}

\begin{figure}[h]
    \centering
    \includegraphics[width=0.8\linewidth]{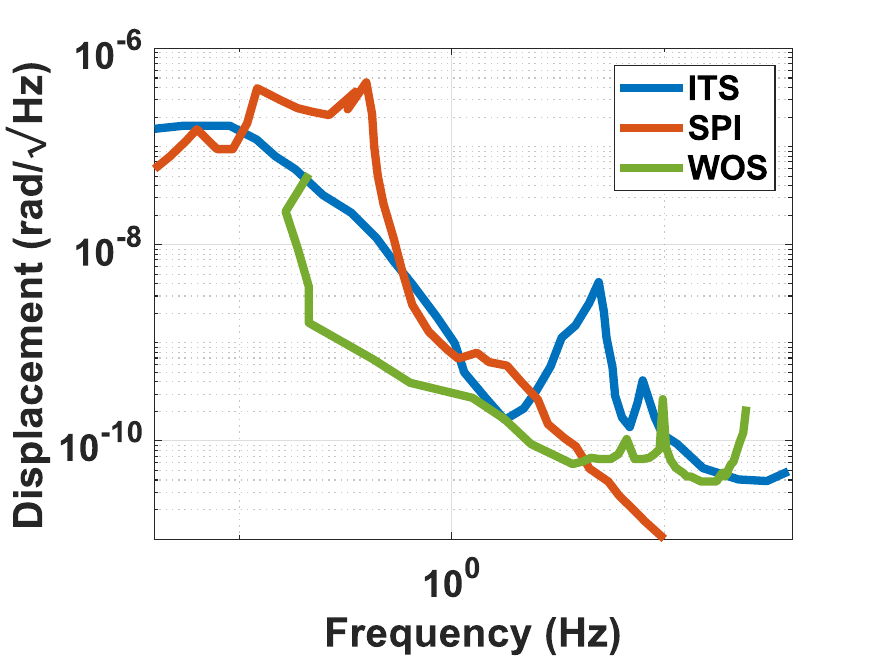}
    \caption{Comparison of the main relative angular sensors recently developed for GW detectors.}
    \label{fig:relative}
\end{figure}

Fig. \ref{fig:relative} compares the performance of the technologies described above with each other. Table \ref{tab:summary} summarizes the technical details of all the technologies described.

\begin{table*}
\centering
\resizebox{0.8\linewidth}{!}{%
\begin{tabular}{c|c|c|c|c}
\rowcolor[HTML]{EFEFEF} 
\textbf{Technology} & \textbf{Type} & \textbf{Resolution}  & \textbf{Frequency at resolution} & \textbf{Size}\\ \hline \hline
OmniSens \cite{omnisens}    & Inertial 6D              & 10$^{-9}$ rad/$\sqrt{Hz}$    & 0.1 Hz  &  600 mm diameter        \\ 
Compact 6D \cite{Ubhi2022}  & Inertial 6D              & 10$^{-9}$ rad/$\sqrt{Hz}$    & 0.1 Hz   & 500 mm diameter         \\ 
CRS\cite{ross2023}          & Inertial rotation sensor & 5 $\times$ 10$^{-12}$ rad/$\sqrt{Hz}$ & 1.5 Hz & 30 cm diameter \\ 
QRS\cite{venka-quar}        & Inertial rotation sensor & 45 $\times$ 10$^{-11}$ rad/$\sqrt{Hz}$ & 1 Hz & 30 cm $\times$ 40 cm \\ 
SILENT\cite{silent}         & Inertial platform        & 2 $\times$ 10$^{-9}$ rad/$\sqrt{Hz}$ & 1 Hz & n/a  \\ 
SPI\cite{Sina2018}          & Interferometric          & 10$^{-9}$ rad/$\sqrt{Hz}$ & 1 Hz & n/a \\ 
WOS\cite{Zhou2001}          & Optical lever            & 10$^{-9}$ rad/$\sqrt{Hz}$ & 1 Hz & 10 cm ca.  \\ 
ALFRA\cite{McCann2021}      & Beam rotation sensor     &  0.1 $\times$ 10$^{-9}$ rad/$\sqrt{Hz}$ & 0.02 Hz & 50 cm $\times$ 50 cm \\ 
BRS\cite{Venkateswara2014}  & Beam rotation sensor     &  0.15 $\times$ 10$^{-9}$ rad/$\sqrt{Hz}$ & 0.1 Hz & ~ 1.5 m \\ 
ITS\cite{Kokeyama2018}      & Interferometric          &  10$^{-10}$ rad/$\sqrt{Hz}$ & 2 Hz & n/a \\ 
\end{tabular}%
}
\caption{Summary of most recent angular motion technologies dedicated to ground-based GW detectors, and their performance.}
\label{tab:summary}
\end{table*}

\subsection{Control strategies}
\label{control}
In this Section, we describe the control strategies used to mitigate tilt effects, and the recent studies for enhancing the currently installed control systems on detectors. We highlight the preferred direction chosen on GW detectors and the proposed strategies for enhancements.\\

As we saw in Sec. \ref{requirements}, detectors and facilities that make use of active isolation adopted strategies to reduce rotational motion on the test masses and keep them more stable. The general ultimate goal is to maintain cavity locking. The main methods involve the use of control feedback systems at cavity suspension points, post-processing tools, or stabilization of the suspension support systems (platforms) to avoid the transmission of the tilt ground motion to the suspension points.We will describe the methods in the following.

\subsubsection{Feedback and feedforward controls}
\begin{figure*}
  
    \includegraphics[width=0.7\linewidth]{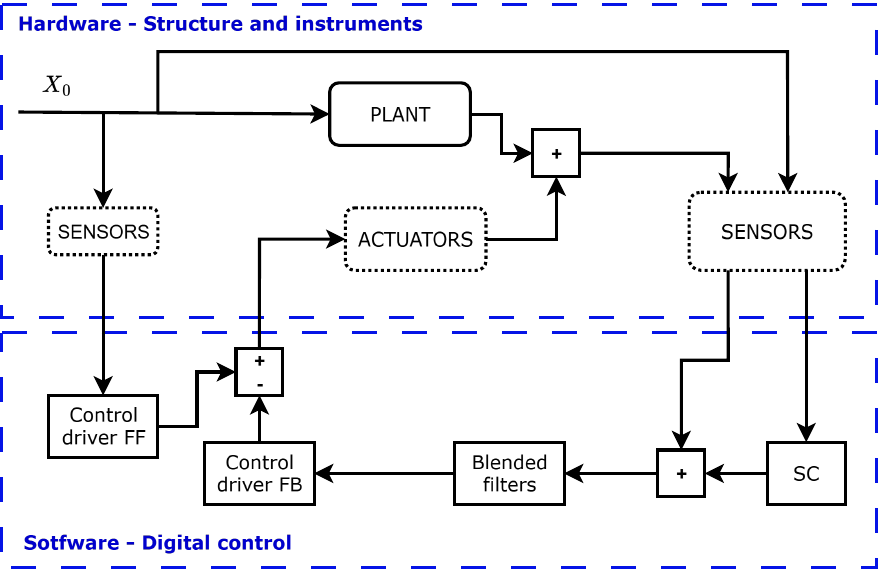}
    \caption{General scheme of a typical LIGO-like control system commonly in use. $X_0$ represents the ground motion. The hardware side includes all the sensors and actuators involved (in all possible degrees of freedom), while the software part is where the digital control happens. The sensor correction (SC) block is shown, and the feedback (FB) and feedforward (FF) schemes are specified. The strategies for reducing tilt motion can be addressed either at hardware level with the use of more or different sensors, or at software level - by changing and improving the filters. Figure adapted from \textit{Matichard et al., 2015} \cite{Matichard2015}.}
    \label{fig:controls}
\end{figure*}

Feedback and feedforward control systems are strategies commonly used on detectors and facilities requiring active isolation. Figure \ref{fig:controls} shows a typical LIGO-like control scheme for seismic motion reduction, adopted by facilities equipped with active platforms (aLIGO, AEI, E-TEST, HOPF), where the main components are highlighted. Generally, these control schemes are composed of multiple sensors and actuators, and multiple control degrees of freedom that use these sensors and actuators. Where possible, diagonalised single input single output (SISO) control structures are  used. The sensor in the SISO control structure is often a super-sensor, a strategy used to account for the general lack of single sensors with adequate performance over the required frequency band \cite{collette2015}. The super-sensor is composed by blending individual sensors such that the sensor with the best signal to noise ratio dominates over the entire frequency range. Sensor diagonalization is normally done in a frequency independent way. However, in case of tilt, frequency dependent diagonalization is required. This is achieved in seismic signals with a sensor correction filter (SC in Fig. \ref{fig:controls}) that subtracts tilt motion at low frequencies, to make diagonalisation of the translation degree of freedom in a frequency dependent manner possible. A feedforward system can be used to contribute to the seismic noise reduction in the translational degrees of freedom, via noise subtraction (see later).\\

\paragraph{\textbf{Applied strategies}}
Reducing the motion of the platforms where the optics are located is a valid action to avoid excessive strain on the cavity locking control systems, because it would reduce the motion transmitted to the mirrors as a natural consequence, allowing for a smaller dynamic range of actuation. This method involves the use of sensors and actuators at various stages of the platforms and can be actuated in two ways, i.e. either improving sensors (as we saw in Sec. \ref{sensing}) or modifying the control systems.\\
\begin{itemize}
    \item In the former way, it is possible to reach impressive levels of suppression. For example, with the SPI technology we showed before the error signal can be suppressed by 3 orders of magnitudes \cite{Sina2018}, while the setup of SILENT can suppress the rotational degrees of freedom of the platform up to $\sim$ 4$\times$ 10$^{-10}$ rad/$\sqrt{\rm Hz}$ at 1 Hz, which is the best measurement achieved for a platform fully isolated by inertial sensors up to now \cite{Lakkis2025}. The research by \textit{Van Dongen et al, 2023} \cite{vanDongen2023} focused on reducing control noise by improving the optical readout of the sensors, specifically by modeling the use of compact interferometers as new, more performative sensors instead of standard shadow sensors, to reduce suspension motion. The model showed that it is possible to reduce the resonant cross-coupling of the suspensions, allowing improved stability for feed-forward control, and resulting in improved sensitivity in the 10–20 Hz band.\\

    \item In the latter way, global optimization algorithms to improve control configurations are among the strategies investigated to stabilize the platform motion, minimize control effort, and reduce the mirror motion at the suspension point stage \cite{Schwartz2020, DiFronzo2025, Carter2020, Qiu2025}. These methods have the advantage to be non-invasive, because usually the installation of a new device is not required. Research by \textit{Schwarz et al, 2020\cite{Schwartz2020}, Di Fronzo et al, 2025 \cite{DiFronzo2025}, Carter et al, 2020\cite{Carter2020}, Tsang et al, 2025\cite{tsang2025}} showed that by optimizing the digital control systems it is possible to achieve better seismic noise reduction and cavity stabilization, with potential positive effects on the differential arm length signal and duty cycle. The work by \textit{Kawazoe et al., 2012\cite{Kawazoe2012}} also showed that improving cavity angular control systems would help surpass the Standard Quantum Limit from 20 Hz above.\\
   Generally, these numerical optimization processes require a cost function, i.e. a function that mimics in the degrees of freedom of interest the implementation of a filter into the control system. Then, the filter's performance is analyzed to estimate its contribution to the platform motion reduction. The cost function must take into account the contributions from each sensor according to their role. Table \ref{tab:contr} shows an example of typical contributions for the construction of a cost function, with an evidence of the tilt contributions from the sensors. Modifying and optimizing the cost functions is currently where the research is focusing to improve detectors seismic noise reduction, without specifically targeting tilt coupling but reducing its effect as a natural consequence.\\

    \item Another branch of research focuses on targeting control issues in presence of unwanted coupling (as in the case of tilt coupling). In control theory, typically, when an input causes the system to initially respond in the opposite direction to its final steady-state output - as in unwanted couplings, such as translation inducing rotation - a phase behavior called \textit{non-minimum phase zero} occurs. This means that the coupling introduces right-half-plane zeros into the system’s transfer functions, which severely limit the achievable bandwidth and robustness of feedback controls. Controlling systems with non-minimum phase zeros are fundamentally more difficult than for minimum-phase systems, because the system initially responds in the opposite direction of the intended motion, introducing phase delays (see details on the level of difficulties of such systems in \textit{Qiu et al., 1993}\cite{qiu1993}). A detailed analytical description of this effect is beyond the scope of this review, and can be found in \textit{Qiu et al., 1993} \cite{qiu1993} and \textit{Skogestad et al., 2005} \cite{controls}. For what concerns a gravitational-wave detector control system, tilt couples in the transfer function system of the inertial sensors measuring the platform motion, actively controlled (as we shown in Sec. \ref{tilt} and in Table \ref{tab:contr}). The strategies usually adopted to minimize the effect are filter designs with specific gain requirements at the target contribution, i.e. near the non-minimum-phase zero frequency, and feed-forward inertial control to reduce disturbances and avoid unstable dynamics \cite{Matichard2015}. This is up to now the consolidated technique adopted on the detectors, although recent research in engineering controls is focusing on how to mitigate the non-minimum-phase zero effect on inertial sensors from both mechanical and digital sides \cite{Lakkis2025, dehaeze2021active}.
\end{itemize}

\begin{table}
\centering
%\resizebox{0.7\textwidth}{!}{%
\begin{tabular}{lll}
\rowcolor[HTML]{EFEFEF} 
\multicolumn{1}{c|}{\cellcolor[HTML]{EFEFEF}\textbf{SENSOR}} & \multicolumn{1}{c|}{\cellcolor[HTML]{EFEFEF}\textbf{Measures}} & \multicolumn{1}{c}{\cellcolor[HTML]{EFEFEF}\textbf{Contributions}} \\ \hline \hline
\multicolumn{1}{l|}{BRS}                                     & \multicolumn{1}{l|}{Ground inertial rotation}                  & $\theta_g$ + n$_{BRS}$                                               \\
\multicolumn{1}{l|}{STS2}                                    & \multicolumn{1}{l|}{Ground inertial translation}               & x$_g$ + n$_STS2$ + $\theta_g\frac{g}{\omega^2}$                                      \\
\multicolumn{1}{l|}{CPS}                                     & \multicolumn{1}{l|}{Relative platform velocity}                & x$_d$ + n$_{CPS}$                                                  \\
\multicolumn{1}{l|}{T240}                                    & \multicolumn{1}{l|}{BSC-platform inertial translation}         & x$_p$ + n$_{T240}$ + $\theta_p\frac{g}{\omega^2}$                                    \\
\multicolumn{1}{l|}{GS13}                                    & \multicolumn{1}{l|}{HAM-platform inertial translation}         & x$_p$ + n$_{GS13}$ + $\theta_p\frac{g}{\omega^2}$                                    \\
   
\end{tabular}%
%}
\caption{Example of the contributions for the construction of a cost function for a LIGO platform control system. The instruments listed are the inertial, relative and rotation sensors currently installed on aLIGO. The tilt contributions from the ground and from the platform are highlighted as $\theta_g$ and $\theta_p$, respectively; $x_{i}$ represents the motion measured, which is the ground (g) in the case of the Streckeisen STS2, the platform (p) for the Trillium T240 and the Geotech GS13, and differential (d = p - g) in the case of the Capacitor Position Sensors (CPS); $n_i$ is the sensor noise associated with each sensor. Table adapted from \textit{Carter et al., 2020} \cite{Carter2020}.}
\label{tab:contr}
\end{table}

\subsubsection{Inertial damping and passive isolation}
Passively isolated detectors generally do not rely on an actuated platform. However, they make use of feedback and feedforward systems locally. In particular, in order to control the mirror motion without injecting noise in the detection band, a technique called \textit{inertial damping} was developed. This strategy makes use of inertial sensors, already present on pendula, and virtual position sensors, to apply feedback force at the higher stage of the pendulum with a wide band high gain. The effect produced is an indirect reduction of tilt motion. The inertial damping control system is illustrated in details by \textit{Losurdo et al., 2001} \cite{losurdo2001inertial} and it it successfully deployed on Virgo. We illustrate in the following how this technique helps reducing ground tilt disturbances.

\begin{itemize}
    \item On Virgo, the IP acts a low pass ﬁlter for seismic noise only along three degrees of freedom: two translations in the horizontal plane and rotation around vertical axis. Ground tilt is then transmitted to the Superattenuator top stage without any attenuation. Moreover the inertial sensors installed on top stage are sensitive to tilt (as we showed earlier) and produce a signal that is misinterpreted as a horizontal acceleration by the control loop, that forces the top stage to move even in absence of horizontal displacement of the ground. Currently, there are no sensors involved for tilt sensing and subtraction, because it was demonstrated that Virgo is not affected by alignment control noise \cite{bader2021} to a level where ground tilt subtraction is necessary. However, in high wind conditions, where it was observed that seismic noise grows up to two or three orders of magnitude between 100 mHz and 1 Hz with its maximum between 400 and 500 mHz (micro-seismic peak) \cite{acernese2015advanced}, tilt effects can become problematic, and are mitigated by a global inverted pendula control (GIPC) \cite{trozzophd, GIPC} that include the application of the inertial damping technique. This is in practice performed by position sensors providing a DC–30 mHz control of the SA position (in order to avoid drifts) \cite{virgo2002inertial} and accelerometers allowing a wideband reduction of the noise in the region of the SA resonances (30 mHz to 5 Hz) \cite{virgo2002inertial}, allowing to indirectly reduce the angular motion partially measured by optical levers.  However, adverse weather conditions can affect Virgo duty cycle to a level where a sensor and a feedback control system is now under study for specifically targeting tilt motion at the inverted pendulum stage \cite{acernese2015advanced} and directly subtract it.

    \item KAGRA is also a passively isolated detector, but thanks to its underground location it does not suffer from significant adverse weather conditions. Likewise Virgo, it is not equipped with a sensing and control scheme addressing tilt motion, however the reduction of this noise is accounted for in the DC control signals in the inverted pendula, similarly to Virgo, and at present it is providing an accepted level of suppression. KAGRA makes also use of a damping system to reduce the residual angular fluctuations of the optics \cite{akiyama2019vibration}.\\
    Although not strictly on topic, it is worth mentioning here that the tilt displacement of the test masses of KAGRA induced by thermal expansion (which is of the order of about 100 $\mu$rad) is addressed via a mechanism for remote mass adjustment. Coil magnets are used for fine alignment (several $\mu$rad) \cite{Nishimoto2020}. For these orders of magnitude, optical levers are efficiently measuring the tilt displacement.
\end{itemize}

\subsubsection{Post-processing and noise cancellation}
Another common technique to limit the disturbance induced by couplings of tilt into the instruments (and thus in the controls), is to apply post-processing strategies or correction by direct measurement of tilt \cite{bernauer2020dynamic}. Different techniques are already available for more general geophysical and seismological requirements \cite{sollberger2020seismological}. They have been further optimized and adapted for gravitational-wave detectors coherent noise subtraction, which is applied via filtering, machine learning and specific post-processing algorithms. We will see these strategies in the following.

\begin{itemize}
    \item Filtering the signal is a common technique that involves witness sensors to measure angular jitter or ground tilt, estimating the noise coupling coefficients, and coherently subtracting the contamination from the main interferometer data via feed-forward. Typically, very high performance filters to distinguish the signals that are dominated by translation from those dominated by ground tilts are designed, taking into account that controls run in real time and filters need to operate at a finite rate as a function of frequency \cite{lantz2009requirements}. This is not a recent technique and no significant improvements for tilt motion reduction have been studied on this topic, although optimizations of filters have been regularly implemented on sites.
    
    \item The use of machine learning algorithms for noise reduction and subtraction has appeared in the last decade. However, there is little literature about specific reduction of noise due to tilt motion. The techniques are mostly investigated as generic frameworks to subtract linear, nonlinear, and nonstationary coupling mechanisms \cite{ormiston2020noise, buchli2025improving, cuoco2021enhancing, cuoco2024applications}. Machine learning techniques dedicated to general seismology are under study \cite{meier2019reliable, chen2019improving, khosro2024machine} and in recent years some research is dedicated specifically to applications on GW detectors seismic requirements, expanding and targeting tilt-to-length coupling in cavities of future detectors. For example, for reducing the tilt motion of the optics and keep cavities stable, Angular Sensing and Control (ASC) systems are widely modeled and used in current and future detectors \cite{Andric2021}. However, ASC systems must be carefully tuned to avoid the injection of control noise and potential opto-mechanical instabilities \cite{maggiore2025,liu2018angular}. For this reason, recent research is also directed toward the use of machine learning techniques to align the cavities and prevent mirror motion \cite{qin2025automated, buchli2025improving}. This branch of research does not directly target ground tilt motion, but it focuses on its effects on the mirror motion directly at the suspension point stage. The motion measured and controlled is then the rotational motion of the mirror itself.\\
    Very recent research focusing on the possibility of deploying machine learning and neural networks for seismic motion modeling and subtraction is expanding with promising outcomes \cite{basalaev2025characterizing, roy2025optimised}. This new branch of research could bring together data analysis and instrumentation experts in the next few years to apply foundation models and AI/ML algorithms to enhance noise mitigation and seismic wave reconstruction.
    
    \item Post-processing techniques are typically used on space-based detectors to account for tilt-to-length disturbances. In this case the tilt motion of the interested components of the satellite (optics, test masses) are not due to the ground, however it is worth briefly mentioning this strategy. For space-based detectors like LISA, an extremely efficient sensing and control system is required for laser pointing \cite{Fan2023, Hartig2025}. Research on accurate sensing systems seeks for sensitivities below the picometer/$\sqrt{\rm Hz}$ at 1 Hz \cite{george2023, Xu2024, Nagano2025} and tilt-to-length coupling limit this sensitivity goal. Actuators are also essential elements in point-ahead angle mechanisms and extreme precision is required. Research on nonlinear effects that could lead to unwanted displacements is conducted to achieve nanoradiant-level tilting \cite{Yan2025, Lin2024}. It is worth noticing that the experimental research on this topic needs to take into account the limitations of tests made on ground (since testbeds for space GW research are ground-based), and adequately address them \cite{Bai2025}, which makes the experimental research more challenging. Studies are dedicated to subtraction of tilt-to-length coupling of the LISA arms \cite{Paczkowski2022, Houba2022}. These techniques are based on optical designs optimized to target and suppress tilt-to-length coupling via different imaging systems techniques \cite{chwalla2020optical, Lin2023, Fan2024, Liu2024, Qiu2025, Zhao2025, Chen2025, Wang2024}.
\end{itemize}

\section{Future perspectives}
\label{future}
It is worth mentioning in this review the plans for tilt motion reduction for the next generation GW detectors. Future GW detectors are planned to be both ground- and space-based.\\
Seismic motion suppression remains at the forefront for future ground-based detectors research. A science case was recently proposed to limit the contribution of the Earth ground noise, that is the lunar-ground-based detector (Lunar Gravitational Wave Antenna, LGWA) \cite{harms_lgwa}. Since the ground noise of the Moon would be considerably lower than the terrestrial one, extremely sensitive sensors are required. The research is focusing then on low-noise floor technology, cryogenic compatible \cite{JvH_lgwa}.\\
The Einstein Telescope, as well as KAGRA, is designed to operate underground \cite{Punturo2013, Abe2022}, and mostly passively isolated. This should reduce the contribution from ground motion. However the research is focusing on improving the performance of sensing, actuation and noise cancellation at their lower limits \cite{Harms2022, bader2021, Allocca2021}. Hybrid configurations of passive and active vibration isolation systems are also under study, to reduce the height of the pendula, maintaining the required level of noise suppression \cite{etest}. Recent studies on the Newtonian Noise (NN) degrees of freedom contributions on GW detectors have also demonstrated how the installation of a tilt sensor can help NN modeling and cancellation \cite{harms2016newtonian, coughlin2018implications}. This research is in expansion in the recent years, with the aim to either substitute seismic arrays currently used for NN cancellation for 6DOF sensors, or improving tilt sensors to verify if a rotation sensor is sufficient to estimate NN \cite{coughlin2018implications}. The improvement of rotation sensors capable of precise tilt measurements could also help subtracting tilt arising from other sources, like wind-induced tilt \cite{Venka2017a}. This particular aspect of using rotation sensors to help with other sources of motion noise could have interesting benefits in the general economy of low-frequency noise reduction.\\
While ground-based detectors like Cosmic Explorer \cite{hall2022cosmic}, upgrades of aLIGO \cite{Lantz2023}, Einstein Telescope \cite{ettdr} will still be affected by ground motion, space-based detectors like LISA \cite{jennrich2009lisa} will not have this noise source. This does not imply that the test masses of LISA will be exempted from motion, since as we mentioned earlier there are other reasons why mirrors could move from operational position. Moreover, given the very long arms (scale of 10$^5$ km) laser pointing control becomes crucial.

\section{Conclusive summary}
\label{conclus}
This paper reviewed the last decade of achievements in addressing angular motion on ground-based gravitational-wave detectors. After an overview of the seismic isolation systems used on the detectors and on the most advanced GW facilities, we showed the most recent technologies developed towards a 6 degrees-of-freedom sensing of the seismic motion. We presented the devices planned for testing and/or installation and their sensitivities. The research seeks a sensitivity of a picoradiant/$\sqrt{\rm Hz}$ at or below 1 Hz.\\

In parallel to the improvement of sensors, digital control schemes are also described in their most recent configuration, optimized with the aim of reducing the motion of the cavities and increasing duty cycle and detector sensitivity. Several strategies have been proposed and tested on sites. The strategies adopted aim to address the tilt motion of the platforms and the one of the mirrors via their dedicated control systems. Post-processing techniques are commonly used and in the last decade, the possibility of using algorithms based on machine learning has emerged.\\

We briefly illustrated the plans for the next generation detectors, highlighting that the ground-based ones will have to address the same challenges related to ground noise, while space-based detectors are not affected by it and can focus on optimizing the laser pointing technologies.\\

We can notice from the discussions presented that the recent research is scoping the most advanced and, at the same time, efficient in terms of affordability and ease of deployment, techniques to reduce as much as possible an important problem affecting ground-based GW detectors and that is rooted in the fundamental working principles of technologies on Earth and control engineering. This branch of research is now at the forefront to improve ground-based gravitational-wave detectors sensitivity at lower frequencies, with interesting and promising outcomes expected in the next years from the advent of machine learning and AI tools. Advances in this research will certainly be beneficial in understanding terrestrial seismic motion and gravitational acceleration, and improving earthquake prediction strategies.

% If in two-column mode, this environment will change to single-column format so that long equations can be displayed. 
% Use only when necessary.
%\begin{widetext}
%$$\mbox{put long equation here}$$
%\end{widetext}

% Figures should be put into the text as floats. 
% Use the graphics or graphicx packages (distributed with LaTeX2e).
% See the LaTeX Graphics Companion by Michel Goosens, Sebastian Rahtz, and Frank Mittelbach for examples. 
%
% Here is an example of the general form of a figure:
% Fill in the caption in the braces of the \caption{} command. 
% Put the label that you will use with \ref{} command in the braces of the \label{} command.
%
% \begin{figure}
% \includegraphics{}%
% \caption{\label{}}%
% \end{figure}

% Tables may be be put in the text as floats.
% Here is an example of the general form of a table:
% Fill in the caption in the braces of the \caption{} command. Put the label
% that you will use with \ref{} command in the braces of the \label{} command.
% Insert the column specifiers (l, r, c, d, etc.) in the empty braces of the
% \begin{tabular}{} command.
%
% \begin{table}
% \caption{\label{} }
% \begin{tabular}{}
% \end{tabular}
% \end{table}

% If you have acknowledgments, this puts in the proper section head.
\begin{acknowledgments}
The authors warmly thank Dr. Felix Bernauer for useful insights on recent developments in rotational seismology. The authors are also grateful to Dr. Alessandro Bertolini and Prof. Ryutaro Takahashi for the correspondence about the current status and future plans of Virgo, ET and KAGRA. Thanks to Dr. Fiona Panther for the useful discussions about the applications of machine learning for GW seismology.\\
The authors thank the OzGrav Centre of Excellence for Gravitational-wave Discoveries for funding this research.
\end{acknowledgments}

\section*{Data Availability Statement}
The datasets used and/or analyzed during the current study are available from the corresponding author on reasonable request.

\section*{References}
% Create the reference section using BibTeX:
\nocite{*}
\bibliography{biblio}

\end{document}